\documentclass[useAMS,usenatbib]{mnras}

\usepackage{graphicx}
\usepackage[T1]{fontenc}
\usepackage{aecompl}
\usepackage{amsfonts}
\usepackage[]{amsmath,amssymb}
\usepackage{color}
\usepackage{xspace}
\usepackage{textcomp}
\usepackage[flushleft]{threeparttable}
\usepackage{xcolor,colortbl}
\usepackage[normalem]{ulem}
\usepackage{multirow}

\newcommand{\msun}{M_\odot}
\graphicspath{{figures/}}

\newcommand{\mytab}{\begin{table}[htb]}
\newcommand{\myfig}{\begin{figure}[htbp]}

\def\lsim{ \lower .75ex \hbox{$\sim$} \llap{\raise .27ex \hbox{$<$}} }

\graphicspath{{figures/}}

\title[]{A machine learning approach to infer the accreted stellar mass fractions of central galaxies in the TNG100 simulation 
}

\author[Shi et al.]{Rui~Shi$^{1,2}$\thanks{sherry97@sjtu.edu.cn}, 
Wenting~Wang$^{1,2}$\thanks{corresponding: wenting.wang@sjtu.edu.cn}, 
Zhaozhou~Li$^{1,2,4}$, Jiaxin~Han$^{1,2}$, Jingjing~Shi$^{3}$, \newauthor
Vicente Rodriguez-Gomez$^{5}$, Yingjie Peng$^{6}$, Qingyang Li$^{1,2}$\\
{}$^{1}$ Department of Astronomy, School of Physics and Astronomy, Shanghai Jiao Tong University, Shanghai 200240, China \\
{}$^{2}$ Shanghai Key Laboratory for Particle Physics and Cosmology, Shanghai 200240, China\\
{}$^{3}$ Kavli IPMU (WPI), UTIAS, The University of Tokyo, Kashiwa, Chiba 277-8583, Japan\\
{}$^{4}$ Centre for Astrophysics and Planetary Science, Racah Institute of Physics, The Hebrew University, Jerusalem 91904, Israel\\
{}$^{5}$ Instituto de Radioastronom\'ia y Astrof\'isica, Universidad Nacional Aut\'onoma de M\'exico, Apdo. Postal 72-3, 58089 Morelia, \\
Mexico\\
{}$^{6}$ Kavli Institute for Astronomy and Astrophysics, Peking University, 5 Yiheyuan Road, Beijing 100871, China
}

\begin{document}

\maketitle

\begin{abstract}

We propose a random forest (RF) machine learning approach to determine the accreted stellar mass fractions 
($f_\mathrm{acc}$) of central galaxies, based on various dark matter halo and galaxy features. 
The RF is trained and tested using 2,710 galaxies with stellar mass $\log_{10}M_\ast/\msun>10.16$ from the 
TNG100 simulation. Galaxy size is the most important individual feature when calculated in 3-dimensions, 
which becomes less important after accounting for observational effects. For smaller galaxies, the 
rankings for features related to merger histories increase. When an entire set of halo and galaxy 
features are used, the prediction is almost unbiased, with root-mean-square error (RMSE) of $\sim$0.068. 
A combination of up to three features with different types (galaxy size, merger history and 
morphology) already saturates the power of prediction. If using observable features, the RMSE 
increases to $\sim$0.104, and a combined usage of stellar mass, galaxy size plus galaxy concentration 
achieves similar predictions. 
Lastly, when using galaxy density, velocity and velocity dispersion profiles as features, which 
approximately represent the maximum amount of information extracted from galaxy images and velocity 
maps, the prediction is not improved much. Hence the limiting precision of predicting $f_\mathrm{acc}$ 
is $\sim$0.1 with observables, and the multi-component decomposition of galaxy images should have similar 
or larger uncertainties. If the central black hole mass and the spin parameter of galaxies can be 
accurately measured in future observations, the RMSE is promising to be further decreased by $\sim$20\%.

\end{abstract}   


\section{Introduction}
\label{sec:intro}

In the structure formation paradigm of $\Lambda$CDM, galaxies form by the cooling and 
condensation of gas at centres of dark matter haloes \citep{1978MNRAS.183..341W}. It is 
usually believed that galaxy formation involves two phases, an early rapid formation of 
``in-situ'' stars through gas cooling and a later phase of mass growth of ``ex-situ'' 
stars through accretion of smaller satellite galaxies, which were originally central 
galaxies of smaller dark matter haloes. These smaller haloes and galaxies, after falling 
into larger haloes, become the so-called subhaloes and satellite galaxies. Satellites 
lose their stellar mass under tidal stripping. Stripped stellar materials build up 
the outskirts of central galaxies and are more metal poor than ``in-situ'' stars, 
which form the diffuse light or the extended stellar halo around the central galaxy 
\citep[e.g.][]{2005ApJ...635..931B}. 

Theoretical studies on the formation of extended stellar haloes involve a few 
different approaches, including analytical models \citep[e.g.][]{2007ApJ...666...20P}, 
hydrodynamical simulations \citep[e.g.][]{2010ApJ...725.2312O,2012MNRAS.425..641L,
2014MNRAS.444..237P,2016MNRAS.458.2371R,2019MNRAS.487..318K} and semi-analytical approaches 
of particle painting/tagging method \citep[e.g.][]{2010MNRAS.406..744C}. It is demonstrated 
that the fraction of accreted stellar mass with respect to the total mass of galaxies is 
higher for more massive galaxies and for elliptical galaxies. The maximum amount of intracluster 
component in the most massive dark matter haloes can be six times as large as the in-situ formed 
stellar mass of the central galaxy \citep[e.g.][]{2009ApJ...693..830Y}. In addition, 
\cite{2016MNRAS.458.2371R,Rodriguez-Gomez2017} also demonstrated the dependence of the 
fraction of ex-situ stars on other halo and galaxy properties in Illustris, such as the 
formation time, merger gas fraction, merger lookback time and merger mass ratio.

Observationally, the determination of accreted stellar mass fractions for galaxies often
relies on a multi-component decomposition of galaxy images or surface brightness profiles 
\citep[e.g.][]{2014MNRAS.443.1433D,Green2017}. The inner and outer components of the 
best-fitting model profiles are usually assumed to represent the in-situ and ex-situ 
components. The decomposition also indicates on average larger fractions of ex-situ stars 
in more massive galaxies and in early-type galaxies \citep[e.g][]{2014MNRAS.443.1433D}, in 
general agreement with predictions from numerical simulations.

However, the multi-component fitting to observed surface brightness profiles of galaxies 
might be subject to unknown systematic uncertainties. For example, \citet{Remus2021} tested
multi-component decomposition using mock galaxy images. They reported that the best-fitting 
inner and outer components deviate from the true in-situ and ex-situ components. The fitting 
is in fact more difficult for late-type galaxies, which have rich substructures and breaks in 
their surface brightness profiles. As a result, \cite{2014MNRAS.443.1433D} adopted triple S\'{e}rsic 
models, with the approach of fitting and fixing the inner two S\'{e}rsic components first, 
and then the outer-most component with small S\'{e}rsic index, lower effective intensity and 
larger effective radius than the inner components is further extracted. Besides, for late-type 
galaxies with stellar mass lower than our Milky Way, the fraction of ex-situ stars is typically 
low ($\sim$1 to 10\%) and tends to show a large scatter, which is very sensitive to systematic 
errors such as the quality of sky background subtraction and the extended PSF wings.

Nowadays, machine learning has more and more applications in the field of cosmology and 
galaxy formation. For example, \cite{Man2019} investigated how to use the L-Galaxies semi-analytical 
galaxy catalogue \citep{2015MNRAS.451.2663H} and the random forest machine learning method 
to predict the host halo mass. \cite{Han2019} used Gaussian process regression to study how halo 
bias depends on multiple halo properties. As machine learning techniques generally do not need to 
assume any parametric forms and can easily handle multiple variables when fitting the data, they 
may provide alternative approaches to model the ex-situ stellar mass fractions in different galaxies. 
In this paper, we investigate such a possibility. We first perform a re-investigation on how the 
ex-situ fraction of stars depends on various physical properties/features of galaxies and dark 
matter haloes, by using the TNG100 simulation from the IllustrisTNG Project and the random 
forest machine learning approach. Compared with the traditional binning method, the random forest 
method, by design, considers the dependence of the ex-situ fraction on various halo/galaxy features 
jointly. It enables us to investigate the importance of different halo/galaxy features and determine 
the ex-situ fractions. We then move on to check the scatter or quality in the model prediction. 

We introduce the data used in this paper, including the galaxy sample from TNG100 and various 
halo/galaxy features in Section~\ref{sec:data}. The random forest machine learning method is introduced 
in Section~\ref{sec:methods}. We present the results about the feature importance ranking, learning 
outcome and scatter control in Section~\ref{sec:results}. Discussions and conclusions are made in 
Section~\ref{sec:disc} and Section~\ref{sec:concl}.

\section{Data}
\label{sec:data}


\subsection{The IllustrisTNG simulation}
\label{sec:TNG100}

Throughout this paper, we use data of the TNG100-1 simulation from the IllustrisTNG Project \citep{Marinacci2018,Naiman2018,Nelson2018,Pillepich2018,Springel2018,2019ComAC...6....2N} for our study. The 
IllustrisTNG series of hydrodynamical simulations include comprehensive treatments of various galaxy 
formation and evolution processes, including metal line cooling, star formation and evolution, chemical 
enrichment and gas recycling. The statistical features of the simulated galaxies have been found to be 
in good agreement with numerous observations, including the global luminosity and stellar mass functions, 
galaxy clustering, color distributions and satellite abundance \citep[e.g.][]{2018MNRAS.475..624N,
2018MNRAS.475..676S,Pillepich2018}. 

The TNG100 simulation was carried out with the Planck 2015 $\Lambda$CDM cosmological model with parameters 
$\Omega_\mathrm{m}=0.3089$, $\Omega_\Lambda=0.6911$, $\Omega_\mathrm{b}=0.0486$, $\sigma_8=0.8159$, $n_s=0.9667$, 
and $h=0.6774$ \citep{Planck2015}. A total of 100 snapshots are saved between $z=127$ and $z=0$. Dark matter haloes 
are identified with the friends-of-friends (FoF) algorithm \citep{Davis1985}. In each FoF group, substructures 
including galaxies are identified with the SUBFIND algorithm~\citep{SUBFIND}. TNG100-1 has a periodic box with 
110.7~Mpc on a side that follows the joint evolution of 1,820$^3$ dark matter particles and approximately 
1,820$^3$ baryonic resolution elements (gas cells and stellar particles). Each dark matter particle has a mass 
of $7.46\times10^6 \msun$, while the the baryonic mass resolution is $1.4\times10^6 \msun$. These 
correspond to a resolution limit of $\sim7.46\times10^{8}\msun$ in halo mass down to 100 dark matter particles.
Galaxies with $\log_{10}M_\ast/\msun< 8$ are likely affected by the resolution limit, while the accurate 
determination of morphology, rotation and shape is likely affected for galaxies with 
$\log_{10}M_\ast/\msun< 9$.

The TNG subhalo merger trees are constructed based on the SubLink algorithm \citep[e.g.][]{Srisawat2013,Avila2014,
Lee2014,2015MNRAS.449...49R}. There are two varieties of merger trees: the baryonic merger trees are built by tracking 
only the stellar particles and star-forming gas cells of subhaloes; The dark matter-only merger trees are following 
the dark matter particles exclusively. By tracing the merger trees, the assembly history of each individual galaxy can 
be tracked down, which helps to determine the origin of every star particle, i.e., whether they are formed in-situ or 
ex-situ. 

\subsection{Ex-situ stellar mass fraction}

Explicitly, we use the stellar assembly catalogue provided on the TNG website \citep{2016MNRAS.458.2371R,
Rodriguez-Gomez2017}, which helps us to determine the origin of each star particle. This catalogue is constructed by tracking the 
baryonic merger trees and determining whether a stellar particle was formed outside of the ``main progenitor branch" of a given 
galaxy. If true, it is considered as an ex-situ stellar particle. Otherwise, the stellar particle is tagged as an in-situ stellar 
particle. For a given galaxy, the amount of stellar mass that was accreted outside of the galaxy versus the total amount of stellar 
mass is defined as the ex-situ stellar mass fraction for this galaxy, which we denote as $f_\mathrm{acc}$ throughout the paper. 

\subsection{Sample selection}

In this study, we first use all true halo central galaxies from TNG100-1, with some mild selections to avoid systems 
undergoing major mergers. In this case, all halo and galaxy features are calculated directly from the simulation and based 
on 3-dimensional coordinates. We then move on to use only observable galaxy features calculated in projection. We will also 
consider a galaxy sample selected in a more realistic way to mimic real observations. We introduce the sample selection in 
the following. Several galaxy features are calculated from mock galaxy images. Details about the mock images and halo/galaxy 
features will be introduced later in Section~\ref{sec:mockimage} and \ref{sec:halogalprop}, respectively.

\subsubsection{Central galaxies in TNG100-1}
\label{sec:central}

For our analysis, we first consider all central galaxies from the redshift $z=0$ snapshot of the TNG100-1 simulation and 
with stellar mass $M_\ast>10^{10.16}\msun$ ($M_\ast>10^{10}\msun$/h). We choose not to push down to galaxies smaller 
than this threshold, whose fraction of accreted stellar mass might be affected by the resolution limit. Besides, in order 
to select relatively isolated systems not perturbed by on-going major merger events, we further require these central 
galaxies to be at least 0.5 magnitudes brighter than their bound satellites in $r$-band, but as we have explicitly 
checked, our results remain very similar if without this selection. In addition, we select galaxies that have faithful 
mock images (see Section~\ref{sec:mockimage} for details).  We finally choose 2,710 central galaxies at $z=0$. We will 
use them to investigate the correlation between $f_\mathrm{acc}$ and their halo/galaxy features. 

For part of our analysis, the sample will be split into two subsamples according to their stellar mass, i.e., 
$\log_{10}M_\ast/\msun>11.20$ and $10.16<\log_{10}M_\ast/\msun \leqslant 11.20$. Here the chosen division of
$\log_{10}M_\ast/\msun=11.20$ is to ensure similar ranges in $\log_{10}M_\ast$ and $f_\mathrm{acc}$.
This is because, global mass and size features as well as $f_{\rm acc}$ itself are highly correlated with the stellar mass. When $\Delta \log_{10}M_\ast$ is narrower for one subsample, $f_{\rm acc}$ as a result will be constrained in a narrow range.
Without significant variations in the sample, the importance rakings of these quantities would be much reduced, resulting in unfair comparisons between the two subsamples. 
The high and low-mass subsamples have 241 and 2,469 galaxies, respectively.

The data sample will be split into two subsets: 1) training set, a subset to train the model; and 2) test set, a 
subset to test the trained model. Throughout the analysis of this paper, we use $70\%$ randomly selected objects 
from the parent sample for training, and the remaining $30\%$ for the test. The readers can check Section~\ref{sec:RF}
for more details.

\subsubsection{Isolated central galaxies}
\label{sec:icg}

It is difficult to directly identify central galaxies in observation. Therefore they are often selected through 
empirical approaches. For example, the build-up of galaxy cluster/group catalogues is based on clustering algorithms
\citep[e.g.][]{2007ApJ...671..153Y,2021ApJ...909..143Y} and color selections for red sequence objects 
\citep[e.g.][]{2014ApJ...785..104R}. Besides, central galaxies are usually brighter than their satellites, and 
thus galaxies that are the brightest within a certain volume in redshift space can be selected to represent 
centrals \citep[e.g.][]{2012MNRAS.424.2574W,2021MNRAS.500.3776W,2021ApJ...919...25W}, which are often referred 
to as locally brightest galaxies, isolated galaxies or isolated central galaxies. All of the approaches, however, 
would lead to different levels of incompleteness and contamination by satellites (impurity). 

To test whether the data incompleteness and impurity might affect the results, we also select a sample of isolated 
central galaxies to test our learning outcome. Specifically, we choose galaxies that are at least 0.5 magnitudes brighter 
in $r$-band than all other companions within the virial radius ($R_{200}$) in projection, and within three times the 
virial velocity ($3V_{\mathrm{vir}}$) along the line of sight from the $z=0$ snapshot of TNG100-1. In addition, we also 
require that the selected galaxies cannot be within the volume defined in the same way around more massive galaxies. We 
choose the $z$-axis of the simulation as the line-of-sight direction. In particular, the virial radius and velocity are 
calculated using the abundance matching formula of \cite{2010MNRAS.404.1111G}. The magnitudes we use for selection 
are the absolute magnitudes from \cite{Nelson2018}, with the inclusion of galaxy dust attenuation and solar neighbourhood 
extinction. The filter response curves are the same as the Sloan Digital Sky Survey \citep[SDSS;][]{Abazajian2009}.
After excluding galaxies which do not have faithful mock images (see Section~\ref{sec:mockimage}), we have 2,231 isolated 
central galaxies from TNG100-1. The purity of the mock galaxy catalogue is as high as $93.49\%$, and the completeness 
fraction is 75.22\% for galaxies with $\log_{10}M_\ast/\msun>10.16$.

\subsection{Mock imaging data}
\label{sec:mockimage}

We will use the SKIRT Synthetic Imaging data for TNG galaxies \citep{Rodriguez-Gomez2019} to calculate a few observable 
galaxy features in projection. The features will be explained in detail in Section~\ref{sec:halogalprop}, and here we only 
briefly introduce the mock images. The mock images are produced by choosing the $z$-axis as the line-of-sight direction.
\citet{Rodriguez-Gomez2019} used SKIRT radiative transfer code \citep{2011ApJS..196...22B,2015A&C.....9...20C} to generate
synthetic images of $\sim$12,500 galaxies for IllustrisTNG, and it is designed to match SDSS observations of
$\mathrm{log_{10}}M_{\ast}/M{\odot}>9.5$ galaxies at $z=0$. Specifically, the mock images were created by assuming that 
the galaxies are located at $z=0.0485$, and have included the effects of dust attenuation and scattering as the photons 
pass through the Interstellar medium (ISM). More details can be found in \citet{Rodriguez-Gomez2019}. 

In our analysis, we use the \texttt{statmorph} package\footnote{https://statmorph.readthedocs.io} to process the mock 
images. We also add the PSF and a simplified background sky model to mimic SDSS observations. The size of the PSF model 
for convolution is simply an azimuthally symmetric Gaussian function with full width at half-maximum (FWHM) being 1.32 arcsec 
in $r$-band, which is the averaged value in SDSS $r$ filter. The pixel size is 0.396 arcsec and the read in each pixel is 
in unit of $e^{-}{\rm s}^{-1}$ instead of nanomaggies. The value of the sky background noise we added is simply
$\sigma_{\mathrm{sky}}=1/10\ e^{-} \, {\rm s}^{-1} \, {\rm pixel}^{-1}$ as recommended on the TNG website. Here we assume 
that all pixels which are $1.5\sigma$ above the sky noise level belong to the source, which is detected by the 
\texttt{photutils} photometry package\footnote{https://photutils.readthedocs.io}. 

The centroid of the galaxy is defined as:
\begin{equation}
    C_{\mathrm{x}} = \frac{\sum\nolimits_{i,j}I_{ij}x_{j}}{\sum\nolimits_{i,j}I_{ij}};\     
    C_{\mathrm{y}} = \frac{\sum\nolimits_{i,j}I_{ij}y_{i}}{\sum\nolimits_{i,j}I_{ij}},
\end{equation}
where $I_{ij}$ is the pixel value of the SKIRT Synthetic image at i-th row and j-th column. $(x_j,y_i)$ is the coordinate of 
the pixel centre. In particular, we only choose those galaxies that have reliable morphological measurements (a quality flag 
equals to zero) and that have mean signal-to-noise level ($S/N$) per pixel larger than 2.5. 

Based on the mock images, we can calculate size and morphological features in projection (see Section~\ref{sec:galaxy_properties} 
for details).

\subsection{Halo and galaxy features}
\label{sec:halogalprop}

In the following, we introduce the halo and galaxy features used in our study. The features are going to be used  
in the random forest method to predict $f_\mathrm{acc}$. In our analysis, we first use the features directly calculated 
from the simulation to do theoretical studies. We then only use the observable features and calculate them in similar 
ways as in real observations. 

\subsubsection{Halo features}
\begin{itemize}
\item $M_{\mathrm{halo}}$

Total mass of all member particles which are bound to the main halo.

\item $R_{200}$

The virial radius of the host dark matter halo, defined as the radius of a sphere centred on 
the potential minimum of this halo, within which the mean matter density is 200 times the 
critical density of the universe, at the time when the halo is considered.

\item $M_{200}$ 

The virial mass of the host dark matter halo, which is defined as the total mass enclosed within $R_{200}$.

\item $\sigma_{v,\mathrm{halo}}$

The total velocity dispersion for all bound member particles of the main halo, which is defined 
as  $\sigma_{v,\mathrm{halo}}=\sqrt{(\sigma_{x}^2+\sigma_{y}^2+\sigma_{z}^2)/3}$. Here 
$\sigma_{x}$, $\sigma_{y}$ and $\sigma_{z}$ are the velocity dispersions of all bound particles 
in the main halo along $x$, $y$ and $z$-axis, respectively.

\item $J_\mathrm{halo}$

The total specific angular momentum of a dark matter halo, defined as
${J_\mathrm{halo}} = \sqrt{{J_x}^2+{J_y}^2+{J_z}^2}$. $J_{x,y,z}$ refer to 
the specific angular momentum along $x$, $y$ and $z$-axis, which are based on all 
bound particles and are in unit of kpc~km/s.

\item $r_\mathrm{half,halo}$

The radius containing half of the total mass of all bound particles in 3-dimensional coordinates.

\item $z_\mathrm{form}$

The redshift when the main halo reaches half of its $z=0$ $M_{\mathrm{halo}}$. 

\item $V_\mathrm{max,halo}$

Maximum velocity of the spherically-averaged rotation curve.

\end{itemize}

\subsubsection{Galaxy features}
\label{sec:galaxy_properties}

\begin{itemize}
\item $\sigma_{v,\ast}$

Stellar velocity dispersion, which is defined as
$\sigma_{v,\ast} = \sqrt{(\sigma_{x,\ast}^2+\sigma_{y,\ast}^2+\sigma_{z,\ast}^2)/3}$, where $\sigma_{x,\ast}$,
$\sigma_{y,\ast}$ and $\sigma_{z,\ast}$ are the velocity dispersions of all star particles belonging to the 
central galaxy and along $x$, $y$ and $z$-axis, respectively. 

To mimic real observations, we choose $\sigma_{z}$ as the line-of-sight velocity dispersion.

\item $M_\ast$

The total stellar mass of the central galaxy. 

In particular, when we use it as an observable feature, we use the projected stellar mass within 
30~kpc from the galaxy centre. The size of 30~kpc is empirically chosen to approximate twice
the Petrosian radius\footnote{Petrosian radius is the circular radius at which the local 
surface brightness $\mu(r)$ equals to $20\%$ of the mean surface brightness enclosed $\mu(<r)$
\citep{Blanton2001}. Twice the Petrosian radius is large enough to contain most of the flux of
galaxies, but would still miss the flux in outskirts of massive galaxies 
\citep[e.g.][]{2013ApJ...773...37H}.} in \cite{Nelson2018}, which agrees 
better with the stellar mass function in SDSS at the massive end.

\item $M_i$ and $M_r$

$M_i$ and $M_r$ are the absolute magnitudes of galaxies in SDSS $i$ and $r$-filters. 

When used as observable features, the magnitudes are defined through the flux within 30~kpc 
to the galaxy centre in projection, with the galaxy dust attenuation and solar neighbourhood 
extinction included. More details can be found in \cite{Nelson2018}. 

\item $g-r$

Rest-frame $g-r$ colour of each galaxy. 

When it is used as an observable feature, dust attenuation and 
solar neighbourhood extinction are included, and the $g$, $r$-bands magnitudes are the flux within 
30~kpc to the galaxy centre in projection. 

\item $r_{90,\mathrm{3D/2D}}$ and $r_{50,\mathrm{3D/2D}}$ 

For theoretical analysis, $r_{90,\mathrm{3D}}$ and $r_{50,\mathrm{3D}}$ are defined as the 3-dimensional radii 
containing 90\% and 50\% of the stellar mass for each galaxy.

When they are used as observable features, $r_{90,\mathrm{2D}}$ and $r_{50,\mathrm{2D}}$ are the projected radii 
containing $90\%$ and $50\%$ of the Petrosian flux in $r$-band, calculated using mock images (see Section~\ref{sec:mockimage}). 
The Petrosian flux is defined as the flux contained within twice the Petrosian radius \citep{2002AJ....123..485S}.

\item merger mass ratio($\mu$)

The maximum stellar mass ratio between the central galaxy and its satellites accreted at $0<z<2$.  
To define the ratio, the mass of the satellite is the maximum stellar mass in its history instead 
of the stellar mass at infall, while the stellar mass of the central is defined at the time of 
accretion. Explicitly, we calculate this value based on the Sublink dark-matter only merger trees 
\citep{2015MNRAS.449...49R}. This feature is not directly observable.

\item Stellar age

The age of the galaxy is defined as the average look back time when star particles were born, 
weighted by the stellar mass of the particle.  

The stellar ages inferred from different stellar population synthesis models may have large 
uncertainties, and as the readers will see from the results in Section~\ref{sec:impsim}, the 
importance of stellar age is low, so we do not include it in our list of observable features.

\item concentration($C_\mathrm{3D/2D}$)

We denote the 3-dimensional concentration and the concentration calculated from projected 
galaxy images as $C_\mathrm{3D}$ and $C_\mathrm{2D}$, respectively. They are defined as 
$C_\mathrm{3D}=r_{90,\mathrm{3D}}/r_{50,\mathrm{3D}}$ and $C_\mathrm{2D}=r_{90,\mathrm{2D}}/
r_{50,\mathrm{2D}}$ \citep[e.g.][]{Graham2005,Cheng2011}. It reflects galaxy morphology. 
High concentration galaxies are mostly elliptical galaxies, while low concentration galaxies 
are mostly spiral  galaxies\citep[e.g.][]{1977ApJ...217..406K,2014MNRAS.443.1433D,2019MNRAS.487.1580W}.

\item $\kappa_\mathrm{rot}$

$\kappa_\mathrm{rot}$ is a kinematic morphology quantity, defined as the ratio between the amount 
of kinetic energy for particles with ordered rotation ($K_\mathrm{rot}$) and the total kinetic 
energy ($K$). This feature was introduced in \cite{Sales2012} as
\begin{equation}
    \kappa_{\mathrm{rot}}=\frac{K_\mathrm{rot}}{K}=\frac{1}{K}\sum\limits_{i}\frac{1}{2}m_{i}(\frac{j_{z,i}}{R_i})^2,
\end{equation}
where $K$ is the total kinetic energy of the stellar component, $m_i$ represents the mass for each star particle, $j_{z,i}$ 
corresponds to the specific angular momentum projected on the $z$-axis and $R_{i}$ is the distance from the particle to 
the galaxy centre, projected on the $x$-$y$ plane. The centre of each halo is defined at the potential minimum. As discussed 
in \cite{Rodriguez-Gomez2017}, $\kappa_\mathrm{rot}$ is a good proxy to the amount of rotation support and is a good measure 
of galaxy morphology.

\end{itemize}

In this study, features related to satellite galaxies are not incorporated in our analysis. This is partly 
because satellites are affected by the resolution limit of the simulation, and about one-third of central 
galaxies used in our analysis do not have any satellites more massive than $10^8\msun$. However, we do have 
explicitly tested to include the total stellar mass in satellites more massive than $10^8\msun$, magnitude 
gaps between centrals and the brightest satellites and the overdensity based on companion counts between 1 
and 2~Mpc to the galaxy centre as features. We found relatively lower importances for them compared with 
other features discussed above, with little improvements in the quality of the learning outcome. Thus we did 
not include them in our analysis. Features related to the gas component and to the metallicity are not used 
either. They are not important for massive galaxies, while tend to have some mild importances for the low-mass 
sample, but the inclusion of them brings very little improvements in the learning outcome.

\section{Methodology}
\label{sec:methods}

\subsection{Decision Trees}
\label{sec:decisiontree}

The decision tree machine learning method \citep{Breiman1983}, which is frequently used in data mining, can mimic 
the process of making a decision and predict the value of a target variable based on several input features. 
Decision trees can be separated into two types: classification trees and regression trees. Classification trees 
are used to determine which class does the target belongs to. Regression trees are used to predict the value of 
the target variable. Our work only involves the regression tree.

The decision tree is essentially a binary partition of the input sample. It starts from the root node and divides 
the input sample into two subsamples according to whether $f_i \leqslant \theta_k$ or $f_i > \theta_k$. Here
$f_i$ is the $i$-th feature of the input sample and $\theta_k$ is a chosen value of $f_i$ adopted for the division.
After this, the root node moves down to two child nodes. The decision tree is then further subdivided according to
different features and values until the size of the subdivided sample on the leaf node is less than
$n_\mathrm{leaf,min}$, a user-specified hyperparameter characterising the minimum subsample size on a leaf node 
for the decision tree model. 

So for each node of the tree, it has to choose the feature $f_i$ and the corresponding feature value $\theta_k$ 
used for sample division. 
This is achieved by maximising the Information Gain, which is defined as

\begin{align}
    I(f_i,\theta_k)=&\mathrm{MSE}_\mathrm{parent\ node}
    -\frac{n_{f_i \leqslant \theta_k}}{n_\mathrm{parent\ node}}\mathrm{MSE}_{f_i \leqslant \theta_k}\nonumber\\
    &-\frac{n_{f_i > \theta_k}}{n_\mathrm{parent\ node}}\mathrm{MSE}_{f_i > \theta_k}, 
\label{eqn:infogain}
\end{align}
where the $\mathrm{MSE}$ is the sample variance of a node
\begin{equation}
    \mathrm{MSE}_\mathrm{node} = \frac{1}{n_\mathrm{node}} \sum\limits_{i\in\mathrm{node}}(y_i-\left\langle{y_{\mathrm{node}}}\right\rangle)^2,
\end{equation}
and $n_\mathrm{parent\ node}$, $n_{f_i \leqslant \theta_k}$ and $n_{f_i > \theta_k}$ are the sample 
size of a parent node or sub-nodes. $n_\mathrm{node}$ represents the sample size of a node. $y$ is the target
variable, and in our case $y$ is the ex-situ fraction in stellar mass, $f_\mathrm{acc}$.
$\left\langle{y_{\mathrm{node}}}\right\rangle$ is the average value of $y$ in the node.

Explicitly, on each node, $f_i$ and $\theta_k$ with the largest Information Gain on this node 
are chosen for the division. When the tree moves down, one feature which has been used on the parent nodes 
can still be used repeatedly on child nodes \footnote{Note the number of features, $K$, to be iterated on 
each node, can be a random subset of the total number of actual features ($N$). Thus $K$ can be smaller than
$N$. $K$ is a user-specified hyperparameter of the decision tree model. In our analysis throughout this paper, 
we choose $K=N$.}. After the tree is constructed, the importance of each feature is defined as the sum of its 
Information Gain over those nodes which used this feature for the division, normalized by the Information 
Gain of all nodes. 

\subsection{Random Forest Method}
\label{sec:RF}

The Random Forest (RF) is an ensemble learning method that constructs multiple decision trees and produces an 
average of the predictions made by each tree in the forest \citep{Breiman2001}. In particular, it uses the 
bootstrap sampling technique for each tree, which means randomly fetching objects from the training data 
set with replacement and with the same sample size. The RF method is widely used because of its simplicity, 
accuracy and fast prediction. Another advantage of RF is that it enables the calculation of the relative 
importance attributed to the input halo and galaxy features. In addition, compared with a single Decision 
Tree, the RF method can reduce the over-fitting issues because it can train each tree after bootstrap sampling 
individually and generate an average result. The RF method has been widely used for a variety of tasks in 
astronomy, e.g., investigating the relation between halo mass and galaxy group properties \citep[e.g.][]{Man2019}, 
exploring how the magnitude gaps between central and satellite galaxies serve as a secondary proxy to the host 
halo mass \citep[e.g.][]{2022arXiv220315222Z}, estimating the virial mass of galaxy clusters in X-ray
\citep[e.g.][]{Green2017}, evaluating the importance of different cluster properties upon determining the 
dynamical status of galaxy clusters \citep[e.g.][]{2022arXiv220315268L}, galaxy morphology classification 
\citep[e.g.][]{2019MNRAS.486.3702S} and calculating the photometric redshift probability distribution function 
\citep[e.g.][]{Carrasco2013}.

In our work, we import the class RandomForestRegressor from the Python package \texttt{scikit-learn} \citep{scikit-learn} 
to build the RF. For more details about this algorithm, we refer to \citet{Breiman2001}. When we deal with the RF, the 
hyperparameters will be tuned to control the growth of the forest and optimize the performance of the prediction's 
bias and variance. There are two major hyperparameters to be tuned: 1) $n_\mathrm{trees}$: the number of trees in 
the forest; 2) $n_\mathrm{leaf,min}$: the minimum number of data required to form a leaf node (the tree will stop 
splitting below this number). We perform convergence tests to seek the best combination of hyperparameters. We 
find for different choices of $n_\mathrm{leaf,min}$, the accuracy of the model changes very quickly at $n_\mathrm{tree} 
\lesssim 20$, and beyond $n_\mathrm{tree}\sim 20$, the accuracy stays almost as constants, indicating the models 
already converge at $n_\mathrm{tree}\sim20$. As a conservative choice, we choose $n_\mathrm{tree}=200$ . In addition, 
numbers of $n_\mathrm{leaf,min}$ between 5 and 25 lead to accuracies not being very different from each other. 
We choose $n_\mathrm{leaf,min}=10$ when all 3-dimensional halo and galaxy features are used, and
$n_\mathrm{leaf,min}=5$ when only observable features are used.

The accuracy of the model is often evaluated from the \textit{out-of-bag} (OOB) scores defined on the OOB data. The OOB 
data is a random subsample of training data that is left out in each tree. When constructing each tree, about $36.8\%$ of 
the training data are not used after bootstrap sampling. After each tree is trained, the outcome is used to make 
predictions for the OOB sample. The predictions made for the OOB samples of all trees will be combined to calculate
the following $R^2$ score, which is the definition of OOB score
\begin{equation}
    R^2(y,\hat{y})=1-\frac{\sum\limits_{i=1}^{i=n_\mathrm{s}}{(y_i-\hat{y_i})^2}}{\sum\limits_{i=1}^{i=n_\mathrm{s}}{(y_i-\bar{y_i})^2}},
    \label{eqn:r2score}
\end{equation}
where $y_i$ and $\hat{y_i}$ are the true and predicted values for each data in the OOB sample, respectively. 
And $\bar{y_i}$ is the mean value. $n_s$ is the size of OOB sample. A larger OOB score means better prediction.

The $R^2$ score can also be calculated for the 30\% test sample similarly and using Equation~\ref{eqn:r2score}, 
which does not depend on the training sample. As we have checked, the $R^2$ score calculated based on either the 
test sample or the OOB sample would not violate the conclusions of this paper. However, the values can vary 
slightly, reflecting the underlying uncertainty due to sample fluctuation. Throughout this paper, we mainly use 
the OOB score for the convergence test to determine the choice of hyperparameters and to estimate the quality 
of leaning outcome, and we use the $R^2$ score based on the test sample to evaluate the importances of individual
features. In addition, we will randomly select 100 different training and test samples, but the fractions of 
training and test samples will be fixed to 70\% and 30\% of the parent sample. The mean score over all test samples
will be used as our final estimate, while the standard deviation of these test samples will be used to estimate 
the uncertainty of the score.

As we will discuss in detail in Section~\ref{sec:impmask}, the default definition of the feature importance (see
Section~\ref{sec:decisiontree}) suffers from the so-called masking effect (see Section~\ref{sec:impmask}), when 
there are strong correlations among different features. Thus we will use a different approach to estimate the 
feature importance ranking. Briefly, we only input one feature to build the RF, and use the $R^2$ score of the 
test sample to quantify the feature importance. We provide more details in Section~\ref{sec:impmask}.

\section{Results}
\label{sec:results}

In this section, we first discuss the correlation among different features. We then present the main results. 
Firstly, we use central galaxies with their 3-dimensional halo and galaxy features (see Sections~\ref{sec:central} 
and \ref{sec:halogalprop}) to investigate the general learning outcome and the feature importance ranking theoretically.
We investigate the importance rankings for both individual features and different feature combinations. We then 
use only observable galaxy features to quantify the learning outcome and their importance rankings. Finally, we 
investigate the level of bias and scatter in the learning outcome, together with some brief discussions on sample 
impurity based on isolated central galaxies.

\subsection{Feature correlations}
\label{sec:impmask}

\begin{figure*} 
\includegraphics[width=1.0\textwidth]{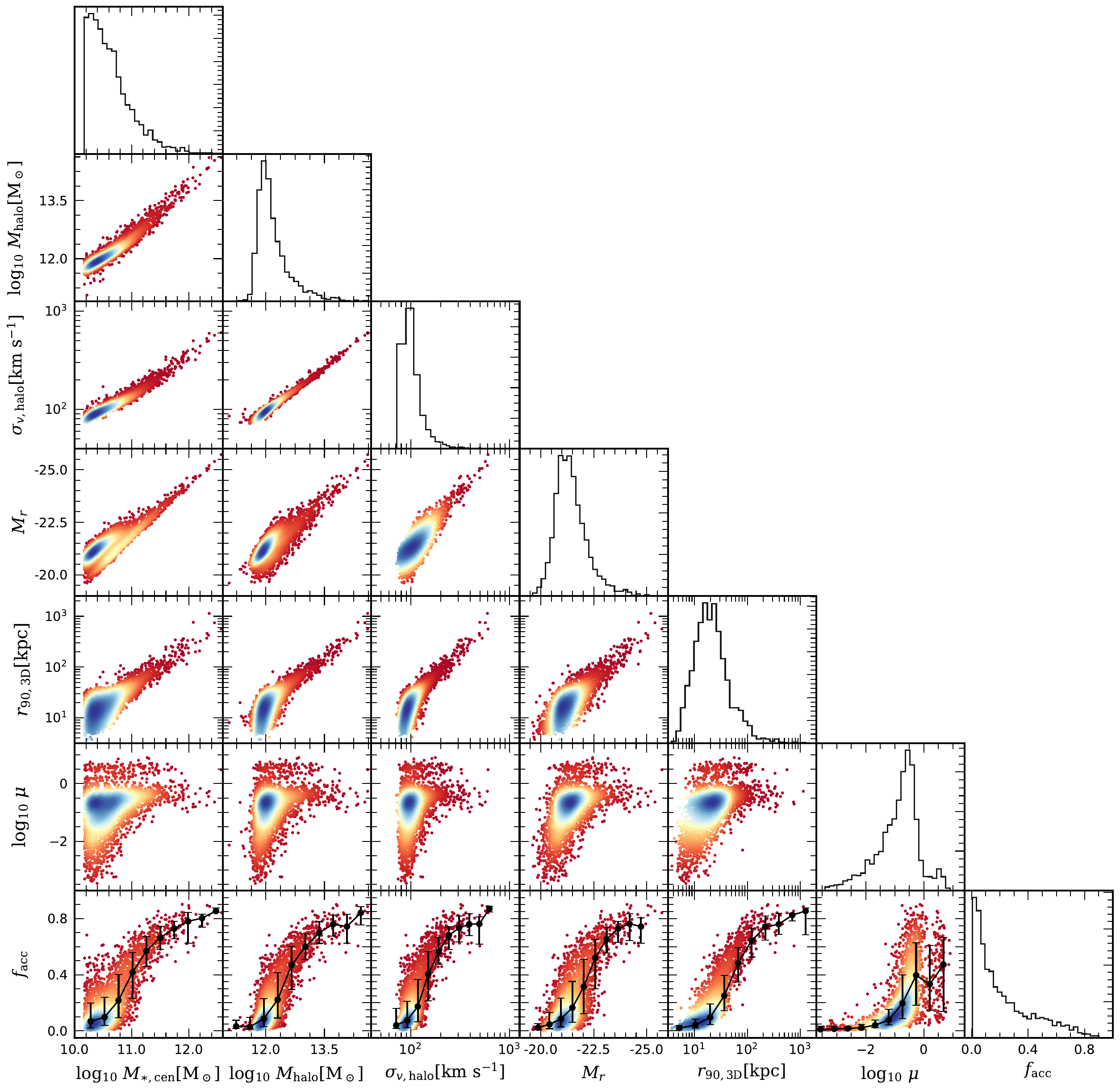}%
\caption{The correlation of a few halo/galaxy features and the ex-situ fraction ($f_\mathrm{acc}$), based on  
central galaxies at $z=0$. The features include $M_{\ast}$, $M_{\mathrm{halo}}$, $\sigma_{v,\mathrm{halo}}$, $M_r$, 
$r_{90,\mathrm{3D}}$ and $\mu$. The readers can find details about these features in Section~\ref{sec:halogalprop}, and 
all the features presented here are calculated in 3-dimensions, rather than in projection. In each panel, the points are 
colour coded according to the number density distributed over the plane formed by the combination of every two different 
features, which are smoothed by Gaussian kernels. The diagonal panels show the probability distribution of each feature. 
The bottom row shows the relationship between $f_\mathrm{acc}$ and these features, in which the black lines 
show the medians of $f_\mathrm{acc}$. The errorbars indicate the 16-th to 84-th percentile ranges centred on 
the medians.
}
\label{fig:correlation figure}
\end{figure*}

The halo and galaxy features used in this paper (see Section~\ref{sec:data}) are not independent of each other.
Figure~\ref{fig:correlation figure} demonstrates the correlation among a few most important features in our 
analysis. The host halo mass of all bound particles ($M_\mathrm{halo}$), the velocity dispersion of the host 
halo ($\sigma_{v,\mathrm{halo}}$), stellar mass ($M_\ast$), the $r$-band absolute magnitudes ($M_r$) of the 
galaxy, and the 3-dimensional radius containing 90\% of the total bound stellar mass ($r_{90,\mathrm{3D}}$) 
are all strongly correlated. The correlations of the above features with the maximum merger mass ratio 
($\mu$)\footnote{A few galaxies have $\mu>1$. This is because the mass of the satellite is defined as the 
maximum stellar mass in its history (see Section~\ref{sec:galaxy_properties} for details). } are weaker 
but still present. 

Fortunately, as we have tested, the learning outcome of the RF method is insensitive to such feature correlations.
This is one of the advantages of the RF method. However, the ranking of feature importance is indeed sensitive to 
such correlations. For example, when both $M_\ast$ and $M_r$ are included, the importance of $M_r$ based on the
default output of the RF method (see Section~\ref{sec:RF}), is only $\sim$1\%. After removing $M_\ast$, the learning
outcome only changes slightly, but the importance of $M_r$ is significantly increased. This is because $M_\ast$ 
already includes most of the information contained in $M_r$. $M_r$ only carries $\sim$~1\% additional information 
than $M_\ast$, while $M_\ast$ perhaps has $>$~1\% additional information than $M_r$. Thus $M_\ast$ always turns out 
to have slightly more Information Gain than $M_r$ for most of the nodes. In other words, the importance of $M_r$ is
suppressed by $M_\ast$, which makes the output feature importance and their rankings challenging to interpret. This 
is referred to as the ``masking effect" of RF \citep{Louppe2014}.

To eliminate such confusions, we try an alternative approach to determine the feature importance. We define the
importance ranking for each feature based on their $R^2$ score ranking (Equation~\ref{eqn:r2score}), obtained by 
only including this feature in the RF training and then applying it to the test sample to calculate the score. 
In this way, the importance of each feature will not be affected by the correlation with other features. In 
addition, we will also investigate the importances of multiple feature combinations in this paper (see 
Sections~\ref{sec:impcomb} and \ref{sec:impobs} for details), and their importance rankings will be represented 
by the $R^2$ score rankings as well, when only including these particular features in the RF model. Notably, 
upon calculating such $R^2$ scores, we have very carefully performed convergence tests to determine the best 
hyperparameters for each feature or feature combination. 

\subsection{Overall learning outcome and individual feature importances of 3-dimensional halo and galaxy features}
\label{sec:impsim}

\begin{figure*} 
\includegraphics[width=0.5\textwidth]{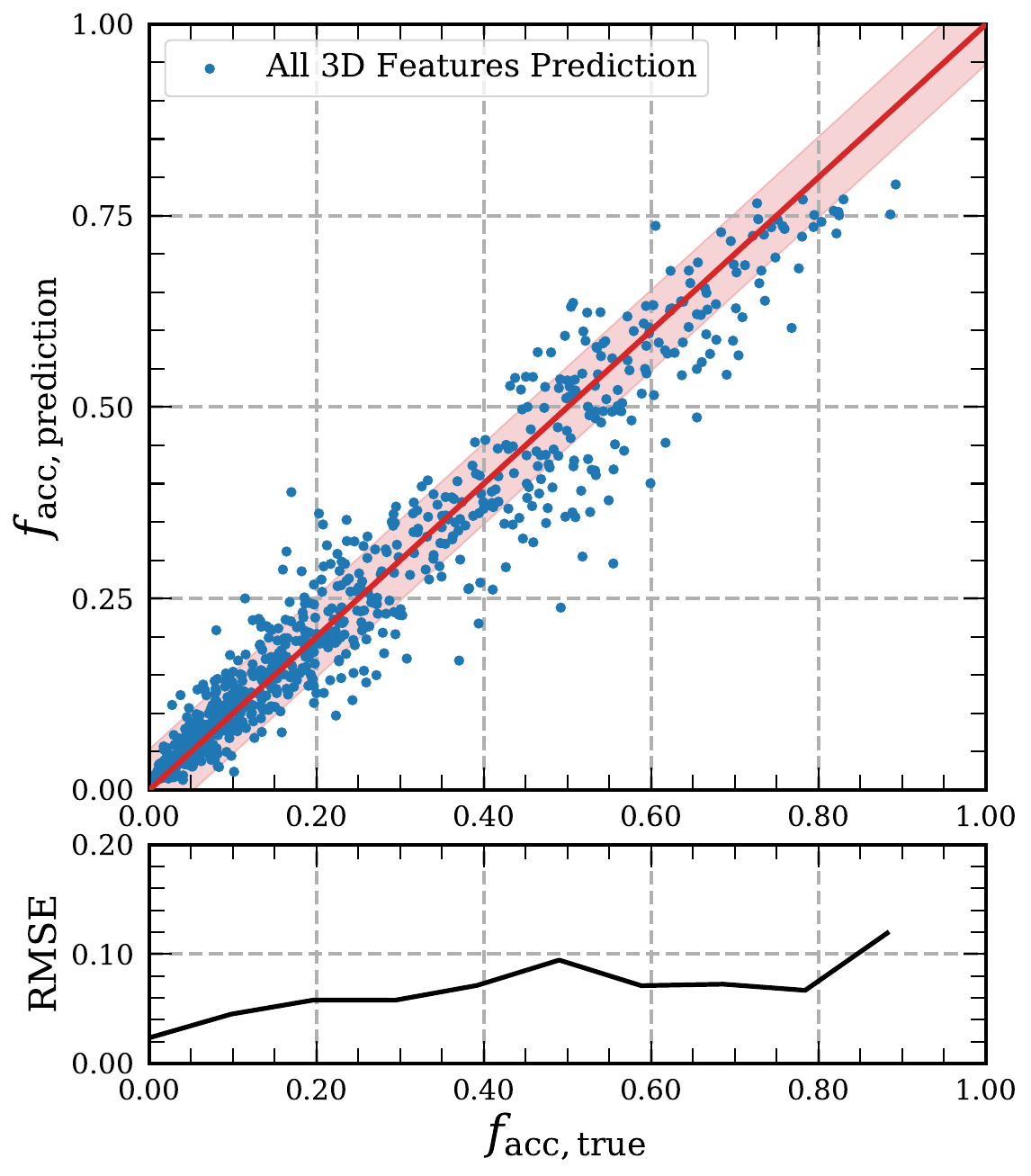}%
\includegraphics[width=0.5\textwidth]{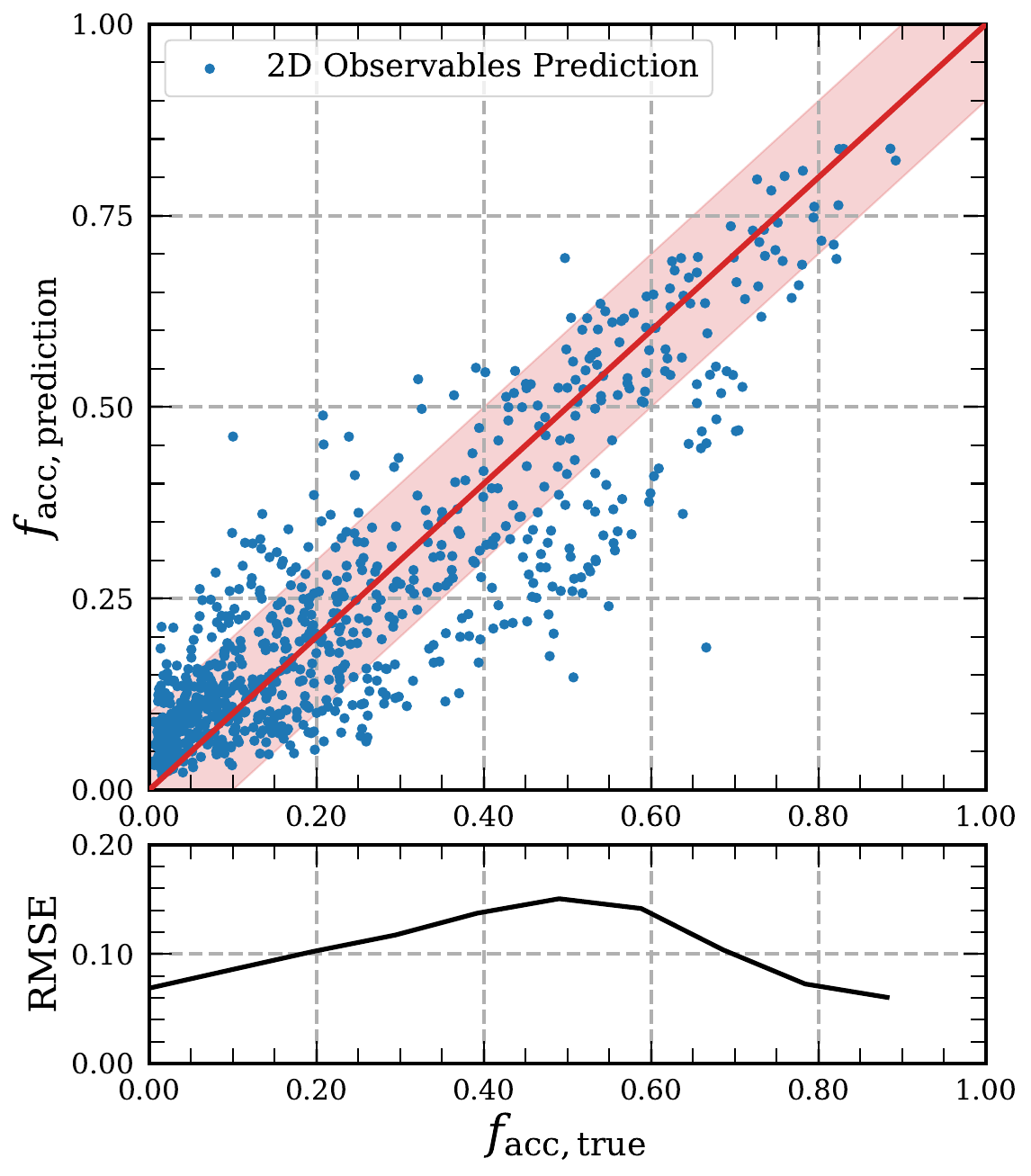}%
\caption{{\bf Left:} The top panel shows the predicted $f_{\mathrm{acc}}$ based on all available features calculated 
in 3-dimensions ($y$-axis), versus the true $f_{\mathrm{acc}}$ from the simulation. This is based on central galaxies 
in TNG100-1, with a mild selection in the magnitude gap with respect to their companions (see Section~\ref{sec:central}). 
The red solid line marks $y=x$, and the red shaded region shows the 1-$\sigma$ standard deviation. The bottom panel 
shows the root-mean-square error, RMSE (Equation~\ref{eqn:RMSE}), as a function of the true $f_{\mathrm{acc}}$.
{\bf Right:} Similar to the left panel, but the prediction is based on only observable galaxy features calculated 
in projection. }
\label{fig:prediction}
\end{figure*}
\begin{table*}
\caption{Feature importances of individual halo/galaxy features. The importance value of each feature is 
represented by the mean $R^2$ score of the test samples, when the model is trained using this particular feature 
(see Section~\ref{sec:impmask} for details). The upper limit of the score is 100\%. The three columns represent 
the full, high and low-mass samples used in our analysis. They are all central galaxies in the simulation, with 
a mild selection in the magnitude gap with respect to their companions (see Section~\ref{sec:central}). Features 
for the full sample are ranked according to the $R^2$ scores, while the integer numbers in the brackets for the 
High-mass and Low-mass columns indicate the ranking, i.e., (1) means the most important feature. The five most 
important individual features are marked in bold fonts. Note the absolute $R^2$ scores of different samples 
are not directly comparable. One can only compare the absolute scores for the same sample, or compare the change in
relative feature importance rankings across different samples.
}
\begin{center}
\begin{tabular}{lccc}\hline
\hline
&&\multicolumn{1}{c}{Importance (\%)}\\
Feature & \multicolumn{1}{c}{Full} & \multicolumn{1}{c}{High-mass}& \multicolumn{1}{c}{Low-mass}\\ \hline
$r_{90,\mathrm{3D}}$ & {\bf 82.12(1)} &{\bf 41.57(1)} & {\bf 70.96(1)}\\
$M_{\mathrm{halo}}$ & {\bf 69.87(2)} & {\bf 36.42(3)} &  {\bf 49.64(2)} \\
$M_{\mathrm{200}}$ & {\bf 69.31(3)} & 34.26(6) & {\bf 48.82(4)}\\
$R_{\mathrm{200}}$ & {\bf 69.31(4)} & 34.25(7) &{\bf 48.82(4)}\\
$M_{\ast}$ & {\bf 67.32(5)} & {\bf 35.52(4)} & 45.08(6) \\
$\sigma_{v,\mathrm{halo}}$ & 66.85(6) & 30.70(10) & 44.78(7) \\ 
$r_\mathrm{half,halo}$ & 64.68(7) & {\bf 34.42(5)} & 41.58(8) \\
$M_i$ & 61.56(8) &  32.95(8) & 36.38(10) \\
$\sigma_{v,\ast}$ & 61.05(9) & 25.65(12) & 35.36(12)\\
$r_{50,\mathrm{3D}}$ & 60.37(10) & {\bf 38.28(2)} &  36.20(11)\\
$J_\mathrm{halo}$ & 59.97(11) & 18.03(15) &  39.70(9) \\
$M_r$ & 58.63(12) & 32.80(9) & 32.72(13) \\
$V_\mathrm{max,halo}$ & 56.73(13) & 28.46(11) &28.08(16) \\
$\mu$ & 47.90(14) & 18.31(14) & {\bf 49.11(3)}\\
$\kappa_\mathrm{rot}$ & 43.06(15) & 24.12(13) & 31.00(14) \\
$C_\mathrm{3D}$ & 36.44(16) & 7.26(17) & 30.00(15) \\
$z_\mathrm{form}$ & 24.15(17) & 15.35(16) & 22.99(17) \\
$g-r$ & 20.46(18) & 0.17(19) & 12.69(18) \\
Stellar age & 17.02(19) & 3.43(18) & 4.95(19) \\
\hline
\label{tbl:imprank}
\end{tabular}
\end{center}
\end{table*}     
\begin{figure} 
\includegraphics[width=0.45\textwidth]{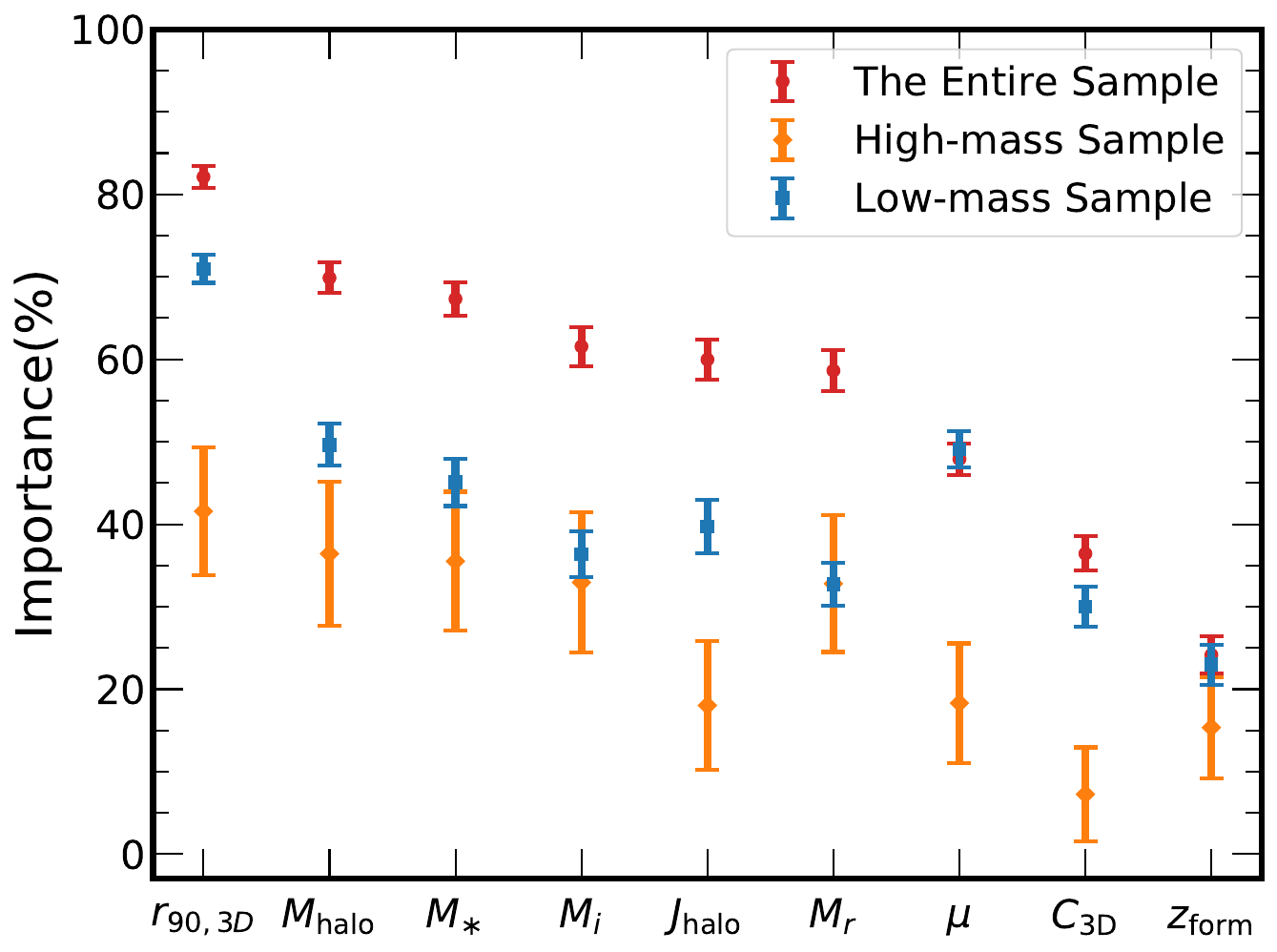}%
\caption{Feature importances of 3-dimensional halo/galaxy features for 
the full (red), high-mass (orange) and low-mass (blue) galaxy samples, respectively. The errorbars are the 
1-$\sigma$ scatters based on randomly dividing the training and test samples 100 times. }
\label{fig:imperr}
\end{figure}

The left panel of Figure~\ref{fig:prediction} shows the predicted values of $f_{\mathrm{acc}}$ versus their 
true values for galaxies in the test sample. The $R^2$ score is 93.94\%. The $f_\mathrm{acc}=f_\mathrm{acc,true}$ 
diagonal line goes well through the data points, indicating no apparent systematic bias. To quantify the scatter, 
we adopt the root-mean-square error (RMSE):
\begin{equation}
    \mathrm{RMSE} = \sqrt{\frac{\sum\limits_{i=1}^{i=n}{\Delta{f_{\mathrm{acc}}^2}}}{n}},
    \label{eqn:RMSE}
\end{equation}
where $\Delta{f_\mathrm{acc}}=f_\mathrm{acc,prediction}-f_\mathrm{acc,true}$, and $n$ is the number of galaxies 
in a given bin of $f_\mathrm{acc, true}$.
The bottom panel of Figure~\ref{fig:prediction} shows the RMSE as a function of $f_\mathrm{acc, true}$. On average, 
RMSE is 0.068. The right panel of Figure~\ref{fig:prediction} is similar, but is based on the prediction 
of using only observable galaxy features, which we will discuss later in Section~\ref{sec:impobs}.

The feature importances for all individual 3-dimensional halo and galaxy features are provided in 
Table~\ref{tbl:imprank}, and for the full, high and low-mass galaxy samples, respectively. For all the 
three samples, $r_{90,\mathrm{3D}}$ is always the most important. This is somewhat surprising, since 
$f_\mathrm{acc}$ is known to be a strong function of stellar mass \citep[][]{2016MNRAS.458.2371R}. However, 
as we have checked, the distribution of $f_\mathrm{acc}$ indeed tends to show the smallest amount of scatter 
in Figure~\ref{fig:correlation figure} when conditioned on $r_{90,\mathrm{3D}}$. This is probably because 
galaxy size is very closely linked to the accretion history. Dry minor mergers tend to build up the outskirts 
of galaxies \citep[e.g.][]{2010ApJ...715..202H,Oser2012,Hilz2013}. Frequent dry minor mergers can cause a 
considerable growth in size with less amount of increase in the mass of galaxies\citep[e.g.][]{Bedorf2013}. 
Observationally, there are also evidences showing that galaxy size is correlated with the environment at 
fixed stellar mass, though still under debates \citep[e.g.][]{2012ApJ...750...93P,2015ApJ...806....3A,2018MNRAS.480..521H,2019A&A...622A..30S}. Here we also 
see $r_{90,\mathrm{3D}}$ is more important than $r_{50,\mathrm{3D}}$, because the accreted materials dominate 
in the outskirts, and thus stars in the outer stellar halo are more strongly correlated with mergers
\citep[e.g.][]{2021ApJ...919..135D}.

For the full sample, a few global features such as halo mass and size ($M_\mathrm{halo}$, $M_\mathrm{200}$ and 
$R_\mathrm{200}$) and the stellar mass of galaxies ($M_\ast$), are among the top five most important individual 
features. Here $M_\mathrm{halo}$ is defined as the total mass of dark matter particles bound to the main subhalo, 
which turns out to be slightly more important than the virial mass, $M_\mathrm{200}$. This probably reflects that 
the physically bound particles are more closely related to the formation of the central galaxy than all particles 
within $R_{200}$. Besides, the halo mass and size features seem to be slightly more important than $M_\ast$. However, 
the typical 1-$\sigma$ uncertainty in the scores is about 2\% (full sample), as shown by Figure~\ref{fig:imperr} 
for a few representative features. Hence the relative ranking for $M_\mathrm{halo}$, $M_\mathrm{200}$, $R_\mathrm{200}$ 
and $M_\ast$ is not very significant compared with the associated uncertainties.

The 6-th and 7-th most important individual features are the total velocity dispersion and the 3-dimensional 
half-mass radius of the host halo, followed by a few stellar features including the $i$-band absolute magnitude ($M_i$), 
the total velocity dispersion of star particles\footnote{We have also checked the velocity dispersion of star particles 
within an aperture of 3 arcsec, by placing galaxies at $z=0.0485$. The central velocity dispersion of galaxies 
has slightly higher importance than the total velocity dispersion, but the difference is not very significant 
compared with the typical uncertainties.} ($\sigma_{v,\ast}$) and the 3-dimensional half-mass radius of stars
($r_{50,\mathrm{3D}}$). The specific angular momentum of the host halo and the maximum circular velocity,
$J_\mathrm{halo}$ and $V_\mathrm{max}$, are less important than $M_\mathrm{halo}$, and the significance is more 
than 1-$\sigma$ (see Figure~\ref{fig:imperr}). $M_r$ is less important than $M_i$, but with a low significance. 
Other features quantifying the assembly history, including the maximum merger mass ratio below $z=2$ ($\mu$), halo
formation time ($z_\mathrm{form}$) and morphology/colour/star formation related features such as $\kappa_\mathrm{rot}$,
$C_{3D}$, $g-r$ and stellar age, are significantly less important than the other global mass and size features.

For the high-mass sample, $r_{50,\mathrm{3D}}$, $M_\mathrm{halo}$, $M_\ast$ and $r_\mathrm{half,halo}$ are 
among the top five most important individual features. Compared with the full sample, the rankings of $M_{200}$ 
and $R_{200}$ now decrease to the 6-th and 7-th. The importances of $\sigma_{v,\mathrm{halo}}$ and $\sigma_{v,\ast}$ 
also decrease, while the rankings of $M_r$, $V_\mathrm{max,halo}$ and $\kappa_\mathrm{rot}$ slightly increase.  
In addition, the few least important features with rankings between 16-th and 19-th in the full sample now still 
have such low importances in the high-mass sample.

For the low-mass sample, the top eight most important individual features are similar to those of the full 
sample, except that $\mu$ now becomes the third most important individual feature, i.e., its ranking is significantly
higher than those in the full and high-mass samples. In addition, the rankings of galaxy concentration,
$C_\mathrm{3D}$, and halo spin, $J_\mathrm{halo}$, increase, but the importances of a few global features such 
as $M_r$, $M_i$, $r_\mathrm{half,halo}$ and $V_\mathrm{max,halo}$ decrease compared with those in the full and
high-mass samples.

On average, the fraction of dark matter is more dominant for low-mass satellite galaxies before infall
\citep[e.g.][]{2010MNRAS.404.1111G}. For a similar amount of growth in stellar mass, the growth in the host dark matter 
halo is stronger for low-mass galaxies, hence a stronger change in the specific angular momentum of the host halo. This 
probably explains why $J_\mathrm{halo}$ has a higher importance ranking than a few other global halo/galaxy features for 
the low-mass sample. 

Interestingly, it seems global halo and stellar features such as the host halo mass, size and stellar mass are significantly 
more important than assembly history related or morphological features for high-mass galaxies, whereas for low-mass 
galaxies, the importances of features related to the assembly histories or to the galaxy morphology ($\mu$, $J_\mathrm{halo}$
and $C_\mathrm{3D}$) have increased.

For both high and low-mass samples, stellar age and $g-r$ colour are not important individually. In fact, we have also 
checked the importances for the star formation rate and specific star formation rate, which are not directly shown in Table~\ref{tbl:imprank}, because they have very low importances individually. This indicates that if only using the 
current star formation activity or colour of galaxies to predict the accreted stellar mass fraction, the predictive power 
is very low.

\subsection{The feature importance of multiple feature combinations}
\label{sec:impcomb}

\begin{figure} 
\includegraphics[width=0.45\textwidth]{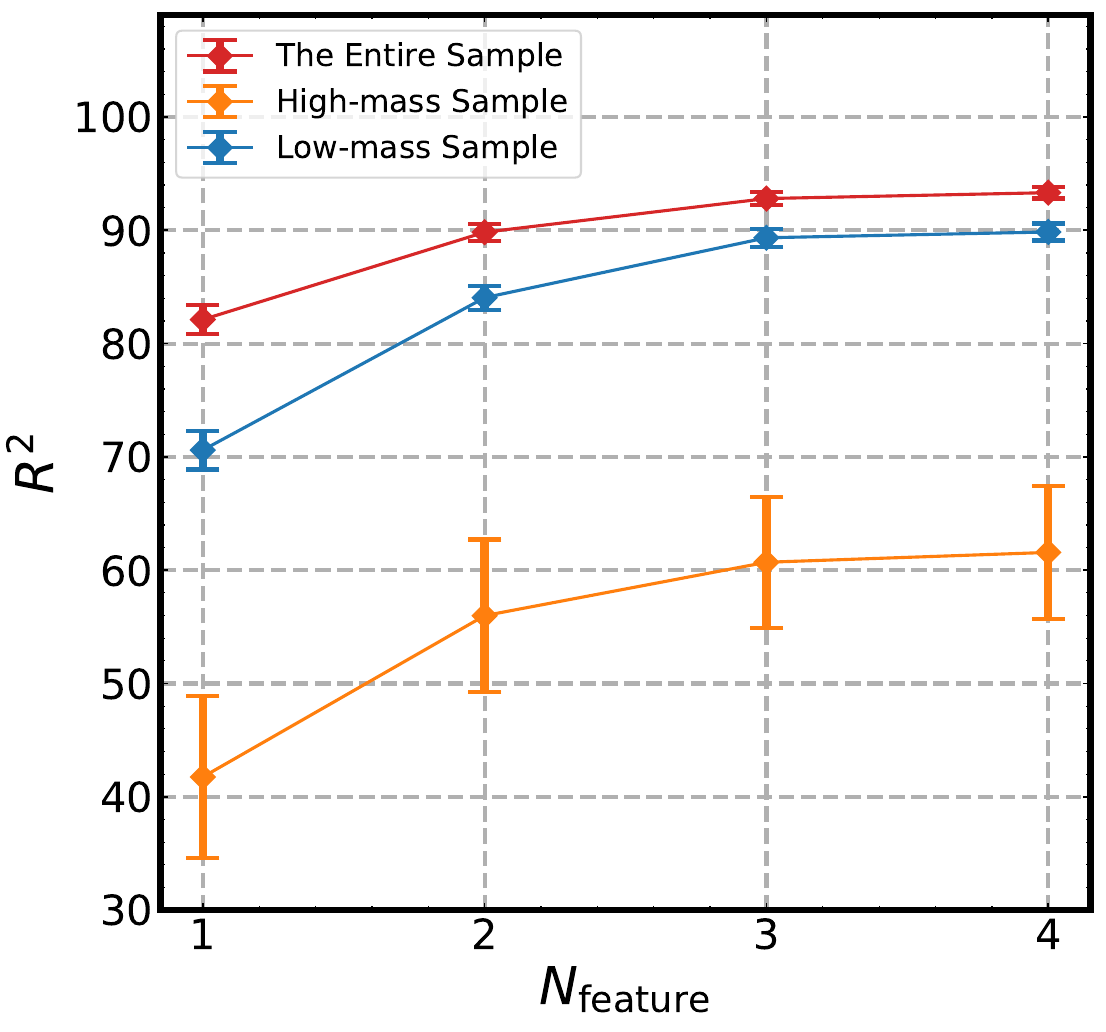}%
\caption{The highest $R^2$ score as a function of the number of features used, for the full (red), 
high-mass (blue) and low-mass (orange) samples, respectively. The $R^2$ scores are presented in percentage. The errorbars represent the 1-$\sigma$ 
scatters of $R^2$ scores. The score saturates at three features. Note the absolute $R^2$ scores of different 
samples are not directly comparable. }
\label{fig:combination}
\end{figure}

\begin{table*}
\caption{The top five highest $R^2$ scores for individual features, two and three feature combinations. This is 
shown for the full, high-mass and low-mass samples separately. The $R^2$ scores are presented in percentage. The
1-$\sigma$ errors in the brackets are given by the scatters among 100 random divisions between the training and test
samples. Note the absolute $R^2$ scores of different samples are not directly comparable. One can only compare the
absolute scores based on the same sample, or compare the change in relative feature importance rankings across 
different samples. }
\begin{center}
\begin{tabular}{ccccccc}\hline
\hline
 Sample & $N_\mathrm{feature}$ & \multicolumn{1}{c}{First} & \multicolumn{1}{c}{Second}& \multicolumn{1}{c}{Third}& \multicolumn{1}{c}{Fourth}& \multicolumn{1}{c}{Fifth}\\ \hline
 \multirow{6}*{Full}
&\multirow{2}*{1}
& $r_{90,\mathrm{3D}}$ & $M_{\mathrm{halo}}$ & $M_{\mathrm{200}}$ & $R_{\mathrm{200}}$  & $M_{\ast}$  \\

 & & 82.12($\pm$1.31) &  69.87($\pm$1.85) & 69.31($\pm$1.93) &69.31($\pm$1.93) & 67.32($\pm$2.00)\\ \cline{2-7}
&\multirow{2}*{2} & $r_{90,\mathrm{3D}}$, $\mu$ & $r_{50,\mathrm{3D}}$, $\kappa_\mathrm{rot}$ & $r_{90,\mathrm{3D}}$, $\kappa_\mathrm{rot}$ & $M_{\mathrm{halo}}$, $\mu$ & $r_{90,\mathrm{3D}}$, $z_\mathrm{form}$\\
&  & 89.83($\pm$0.76) &86.39($\pm$0.91) & 86.36($\pm$1.03) & 85.08($\pm$1.08) & 84.85($\pm$1.12) \\ \cline{2-7}
&\multirow{2}*{3} &  $r_{90,\mathrm{3D}}$, $\mu$, $\kappa_\mathrm{rot}$ & $r_{90,\mathrm{3D}}$, $\mu$, $C_\mathrm{3D}$ & $r_{90,\mathrm{3D}}$, $r_{50,\mathrm{3D}}$, $\mu$ & $r_{50,\mathrm{3D}}$, $\mu$, $C_\mathrm{3D}$ & $r_{50,\mathrm{3D}}$, $\mu$, $\kappa_\mathrm{rot}$ \\
 & & 92.80($\pm$0.58) & 91.55($\pm$0.66) & 91.50($\pm$0.64) & 91.36($\pm$0.59) & 91.17($\pm$0.65)\\ \cline{1-7}
 \multirow{6}*{High-mass}&
 \multirow{2}*{1}
 &$r_{90,\mathrm{3D}}$ & $r_{50,\mathrm{3D}}$ & $M_{\ast}$ & $M_{\mathrm{halo}}$  & $r_\mathrm{half,halo}$  \\

 & & 41.75($\pm$7.12) &  39.67($\pm$6.22) & 35.99($\pm$7.50) &35.94($\pm$7.72) & 34.13($\pm$7.61)\\ \cline{2-7}
&\multirow{2}*{2} & $r_{50,\mathrm{3D}}$, $\mu$ & $r_{90,\mathrm{3D}}$, $\mu$ & $M_{\mathrm{halo}}$, $\mu$& $M_{\ast}$, $\mu$ & $M_{\mathrm{200}}$, $\mu$\\
 & & 55.99($\pm$6.72) &55.23($\pm$6.46) & 51.03($\pm$8.44) & 50.88($\pm$7.07) & 50.38($\pm$8.23) \\ \cline{2-7}
&\multirow{2}*{3} &  $r_{50,\mathrm{3D}}$, $\mu$, $\kappa_\mathrm{rot}$ & $r_{90,\mathrm{3D}}$, $\mu$, $C_\mathrm{3D}$ & $r_{90,\mathrm{3D}}$, $r_{50,\mathrm{3D}}$, $\mu$ & $r_{50,\mathrm{3D}}$, $\mu$, $C_\mathrm{3D}$ & $r_{90,\mathrm{3D}}$, $\mu$, Stellar Age \\
 & & 60.70($\pm$5.79) & 60.55($\pm$6.14) & 59.60($\pm$6.18) & 58.51($\pm$6.80) & 57.64($\pm$6.46)\\ \cline{1-7}
 \multirow{6}*{Low-mass}&
 \multirow{2}*{1}
& $r_{90,\mathrm{3D}}$ & $M_{\mathrm{halo}}$ & $\mu$ & $R_{\mathrm{200}}$  & $M_{\mathrm{200}}$ \\

 & & 70.60($\pm$1.71) &  49.62($\pm$2.70) & 48.93($\pm$2.24) &48.75($\pm$2.74) & 48.74($\pm$2.72)\\ \cline{2-7}
&\multirow{2}*{2} & $r_{90,\mathrm{3D}}$, $\mu$ & $r_{50,\mathrm{3D}}$, $\kappa_\mathrm{rot}$ & $r_{90,\mathrm{3D}}$, $\kappa_\mathrm{rot}$& $M_{\mathrm{halo}}$, $\mu$ & $r_{90,\mathrm{3D}}$, $z_\mathrm{form}$\\
&  & 84.05($\pm$1.05) &78.51($\pm$1.18) & 78.50($\pm$1.18) & 75.87($\pm$1.57) & 75.62($\pm$1.60) \\ \cline{2-7}
&\multirow{2}*{3} &  $r_{90,\mathrm{3D}}$, $\mu$, $\kappa_\mathrm{rot}$ & $r_{90,\mathrm{3D}}$, $\mu$, $C_\mathrm{3D}$& $r_{90,\mathrm{3D}}$, $r_{50,\mathrm{3D}}$, $\mu$ & $r_{50,\mathrm{3D}}$, $\mu$, $C_\mathrm{3D}$ & $r_{50,\mathrm{3D}}$,  $\mu$, $\kappa_\mathrm{rot}$ \\
 & & 89.34($\pm$0.78) & 86.78($\pm$0.97) & 86.74($\pm$0.96) & 86.41($\pm$0.93) & 86.07($\pm$0.89)\\ \hline
\hline
\label{tbl:imp_combine}
\end{tabular}
\end{center}
\end{table*}                  

In Section~\ref{sec:impsim}, we have investigated the importance rankings of different individual features. In Table~\ref{tbl:imprank}, we can see that the highest $R^2$ score of individual features is 82.12\% for the full 
sample, but a higher $R^2$ score of 93.94\% can be achieved when all halo and galaxy features are used. This means
using only one individual feature to predict $f_\mathrm{acc}$ is not enough. However, using a large number of features
to determine $f_\mathrm{acc}$ could be too redundant in practice. In this sense, we try to determine $f_\mathrm{acc}$ 
with the combination of a limited number of features and investigate the importance rankings of these combinations.

To investigate this, we calculated the $R^2$ scores for all possible combinations of every two, three and four 
features, same as done in \citet{2022arXiv220315268L} and \citet{2022arXiv220315222Z}. The highest $R^2$ score with respect to the number of combined features is shown in Figure~\ref{fig:combination}. 
The red, orange and blue curves are for the full, high and low-mass samples, respectively. The errorbars represent the
1-$\sigma$ scatters by dividing the training and test samples for 100 times. We find the curves become flattened beyond a number of three features, indicating a combination of up to three features can already saturate the score of the prediction. The highest $R^2$ score achieved with three features is 92.80\% for the full sample, which is very close to the score when 
all available halo and galaxy features are used (93.94\%). This means a lot of information related to $f_\mathrm{acc}$ can 
be well explained by using only three features in combination. 

Table~\ref{tbl:imp_combine} shows the top five most important features or feature combinations, for the cases of 
individual features, two and three combined features. We show this for the full, high and low-mass samples separately. 
Note although we have ranked the different feature combinations in Table~\ref{tbl:imp_combine}, the significances
in the ranking are not very high compared with the errors ($R^2$ scores and associated errors are shown below the 
feature names). 

We find $r_{90,\mathrm{3D}}$, which is the most important single feature, frequently appears in all different combinations 
as well. For the combination of two features, the five most important cases are often one global mass or size feature ($r_{90,\mathrm{3D}}$, $r_{50,\mathrm{3D}}$, $M_\mathrm{halo}$ or $M_\ast$), plus another assembly history related or 
morphological feature ($C_\mathrm{3D}$, $\mu$, $\kappa_\mathrm{rot}$, $z_\mathrm{form}$ or stellar age). For the case of 
three features, those most important combinations are often one mass/size feature, plus another two assembly/morphological 
features. In fact, despite the fact that $\mu$ is not very important as an individual feature in the full and high-mass 
samples, and $\kappa_\mathrm{rot}$, for example, is not a very important individual feature in the full, high and 
low-mass samples, they frequently appear in the most important two and three feature combinations. This indicates 
that assembly or morphological features carry the most amount of independent information from those in global mass/size
features, and thus after combined together, they become the most important. On the other hand, the few top important 
individual features in Table~\ref{tbl:imprank}, such as $r_{90,\mathrm{3D}}$, $M_\mathrm{halo}$, $M_{200}$, $R_{200}$ 
and $M_\ast$, are highly correlated with each other, carrying redundant information. This explains why the combination of, 
for example, $r_{90,\mathrm{3D}}$ and $\mu$, becomes more important than the combination of $r_{90,\mathrm{3D}}$ and 
$M_\ast$ for instance.

Interestingly, we notice two features which are both not very important individually, may become very important 
after being combined together. For example, the combination of $r_{50,\mathrm{3D}}$ and $\kappa_\mathrm{rot}$ rank 
the second most important for the full sample and the low-mass sample. $r_{50,\mathrm{3D}}$ also appears multiple 
times in the five most important three feature combinations. However, $r_{50,\mathrm{3D}}$ only ranks the 10-th and 
11-th important as an individual feature for the full and low-mass samples. The importance of $\kappa_\mathrm{rot}$ 
as an individual feature is also very low for the full, high and low-mass samples. 

Our results show, very promisingly, one can choose to use one global mass or size feature of the galaxy or 
host halo, in combination with another two features reflecting the assembly history or morphology to predict 
$f_\mathrm{acc}$. In fact, using the combination of $r_{90,\mathrm{3D}}$, $\mu$ and $\kappa_\mathrm{rot}$, 
the RMSE is $\sim$0.072, which is very similar to the RMSE of 0.068 in the left panel of Figure~\ref{fig:prediction},
when all available halo and galaxy features are used. In other words, the inclusion of many other halo and galaxy 
features almost does not help to decrease the RMSE.

In the next subsection, we move on to discuss the RF learning outcome and importance ranking 
when only observable galaxy features are used.

\begin{table}
\caption{Similar to Table~\ref{tbl:imprank}, but shows the individual feature importance rankings for only 
observable galaxy features in projection. The three most important individual features are marked in bold fonts.
Note the absolute $R^2$ scores of different samples are not directly comparable. One can only compare the absolute 
scores for the same sample, or compare the change in relative feature importance rankings across different samples.}
\begin{center}
\begin{tabular}{lccc}\hline
\hline
&&\multicolumn{1}{c}{Importance (\%)}\\
Feature & \multicolumn{1}{c}{Full} & \multicolumn{1}{c}{High-mass} & \multicolumn{1}{c}{Low-mass}\\ \hline
$M_{\ast}$ & {\bf64.71(1)} & {\bf 23.43(2)} & {\bf 48.46(1)} \\
$M_i$ & {\bf 59.34(2)} &  14.82(4) & {\bf 41.79(2)} \\
$M_r$ & {\bf 57.33(3)} & 14.43(6) & {\bf 39.32(3)} \\
$r_{90,\mathrm{2D}}$ & 53.18(4) & {\bf 34.43(1)} & 36.20(4)\\
$\sigma_{z,\ast}$ & 52.59(5) & 14.72(5) & 31.71(6)\\
$C_\mathrm{2D}$ & 35.49(6) & 7.62(7) &32.31(5) \\
$r_{50,\mathrm{2D}}$ & 31.95(7) & {\bf 20.95(3)} & 16.50(7)\\
$g-r$ & 16.56(8) & 7.59(8) & 10.10(8) \\
\hline
\label{tbl:imprankobs}
\end{tabular}
\end{center}
\end{table}      

\subsection{The learning outcome and feature importances of observable galaxy features}
\label{sec:impobs}

\begin{figure} 
\includegraphics[width=0.49\textwidth]{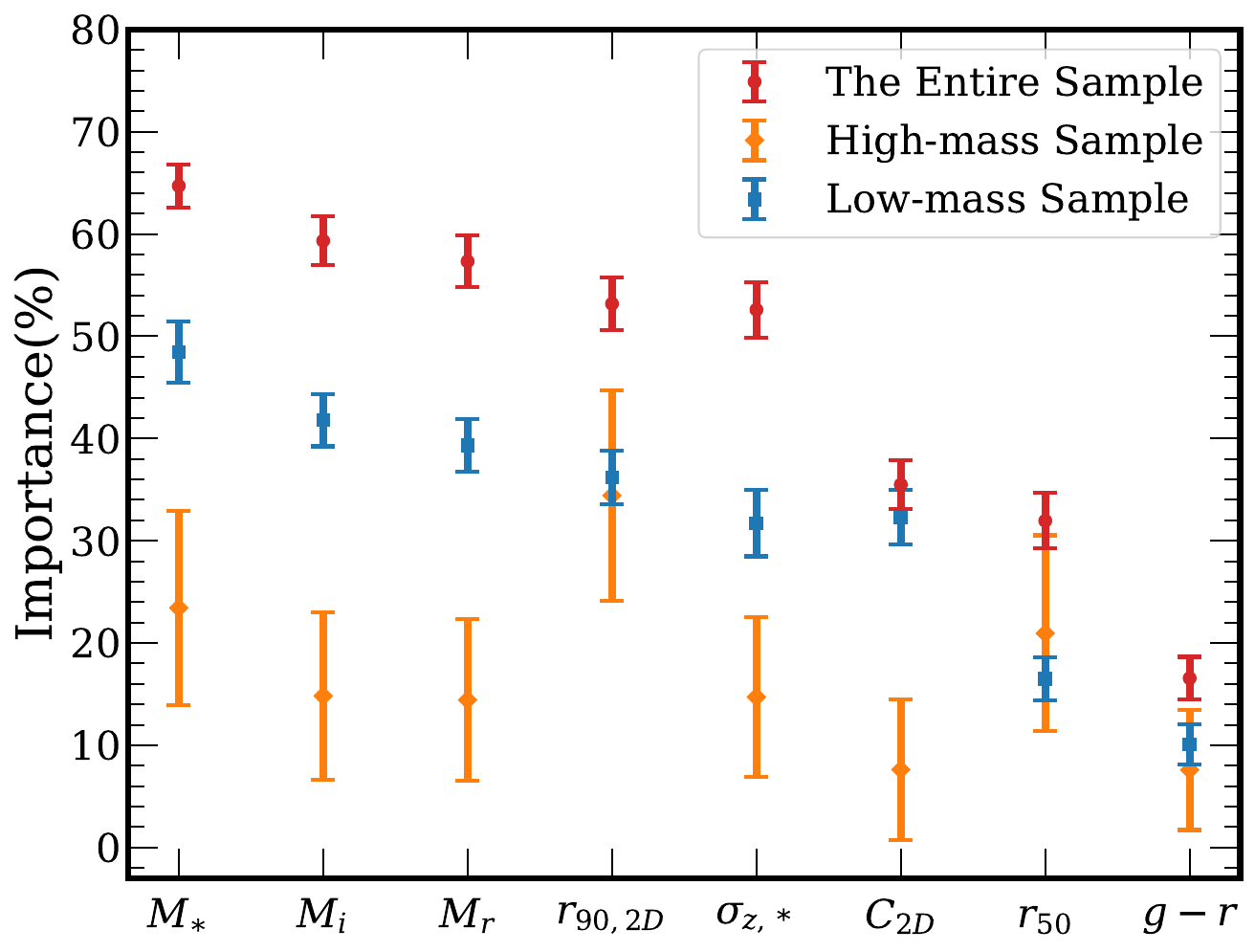}%
\caption{Feature importances of individual observable galaxy features, and for the full (red), 
high-mass (orange) and low-mass (blue) galaxy samples, respectively. Errorbars are the 1-$\sigma$ 
scatters based on randomly dividing the training and test samples 100 times. The larger errorbars 
for the high-mass sample is due to the smaller sample size.
}
\label{fig:impobs}
\end{figure}
\begin{table*}
\caption{The top five most important individual features, two and three feature combinations, and the associated $R^2$ scores, for the full, high and low-mass samples. This is only for observable features. The $R^2$ scores are
presented in percentage. The 1-$\sigma$ scatters in the brackets are given by randomly dividing the training and test
samples for 100 times. Note the absolute $R^2$ scores of different samples are not directly comparable. One can only
compare the absolute scores for the same sample, or compare the change in relative feature importance rankings across
different samples. }
\begin{center}
\begin{tabular}{ccccccc}\hline
\hline
 Sample & $N_\mathrm{feature}$ & \multicolumn{1}{c}{First} & \multicolumn{1}{c}{Second}& \multicolumn{1}{c}{Third}& \multicolumn{1}{c}{Fourth}& \multicolumn{1}{c}{Fifth}\\ \hline
 \multirow{6}*{Full}
&\multirow{2}*{1}
& $M_{\ast}$ & $M_{i}$ & $M_{r}$ & $r_{90,\mathrm{2D}}$ & $\sigma_{z,\mathrm{*}}$    \\

&  & 64.71($\pm$2.11) &  59.34($\pm$2.41) & 57.33($\pm$2.50) &53.18($\pm$2.57) & 52.59($\pm$2.72)\\ \cline{2-7}
&\multirow{2}*{2} & $M_{\ast}$, $r_{90,\mathrm{2D}}$ & $M_{\ast}$, $C_\mathrm{2D}$ & $M_i$, $C_\mathrm{2D}$ & $M_r$, $C_\mathrm{2D}$& $r_{90,\mathrm{2D}}$, $C_\mathrm{2D}$\\
 & & 70.74($\pm$1.91) &68.97($\pm$1.89) & 68.88($\pm$2.01) & 68.73($\pm$2.03) & 68.16($\pm$1.80) \\ \cline{2-7}
&\multirow{2}*{3} &  $M_{\ast}$, $r_{90,\mathrm{2D}}$, $C_\mathrm{2D}$ & $M_{\ast}$, $C_\mathrm{2D}$, $r_{50,\mathrm{2D}}$ & $M_i$, $r_{90,\mathrm{2D}}$, $C_\mathrm{2D}$& $M_r$, $r_{90,\mathrm{2D}}$, $C_\mathrm{2D}$ & $M_i$, $C_\mathrm{2D}$, $r_{50,\mathrm{2D}}$\\
 & & 75.76($\pm$1.66) & 75.36($\pm$1.66) & 74.85($\pm$1.64) & 74.70($\pm$1.69) & 74.16($\pm$1.71)\\ \cline{1-7}
 \multirow{6}*{High-mass}&
 \multirow{2}*{1}
 & $r_{90,\mathrm{2D}}$ &$M_{\ast}$&$r_{50,\mathrm{2D}}$&$M_i$&$\sigma_{z,\ast}$\\
 &&34.43($\pm$10.28)&23.43($\pm$9.50)&20.95($\pm$9.59)&14.82($\pm$8.19)&14.72($\pm$7.82) \\\cline{2-7}
 &\multirow{2}*{2} & $r_{90,\mathrm{2D}}$, $g-r$ & $r_{90,\mathrm{2D}}$, $C_\mathrm{2D}$ & $r_{90,\mathrm{2D}}$, $r_{50,\mathrm{2D}}$& $r_{90,\mathrm{2D}}$, $\sigma_{z,\ast}$& $r_{90,\mathrm{2D}}$, $M_{\ast}$\\
 & & 41.37($\pm$9.37) &39.47($\pm$9.88) & 39.39($\pm$10.69) & 38.04($\pm$10.32) & 37.58($\pm$11.10) \\ \cline{2-7}

&\multirow{2}*{3} &  $r_{90,\mathrm{2D}}$, $\sigma_{z,\ast}$, $g-r$ & $r_{90,\mathrm{2D}}$, $M_{\ast}$, $C_\mathrm{2D}$ & $r_{90,\mathrm{2D}}$, $M_{\ast}$, $g-r$  & $r_{90,\mathrm{2D}}$, $\sigma_{z,\ast}$, $C_\mathrm{2D}$ & $r_{90,\mathrm{2D}}$, $M_r$, $g-r$ \\
 & & 44.23($\pm$9.63) & 43.12($\pm$10.72) & 42.87($\pm$10.87) & 42.80($\pm$10.09) & 42.19($\pm$10.40)\\ \cline{1-7}
 \multirow{6}*{Low-mass}&
 \multirow{2}*{1}
& $M_{\ast}$ & $M_i$ & $M_r$ & $r_{90,\mathrm{2D}}$  & $C_\mathrm{2D}$ \\

 & & 48.46($\pm$3.01) &  41.79($\pm$2.56) & 39.32($\pm$2.56) &36.20($\pm$2.61) & 32.31($\pm$2.65)\\ \cline{2-7}

&\multirow{2}*{2} & $M_{\ast}$, $r_{90,\mathrm{2D}}$ & $r_{90,\mathrm{2D}}$, $C_\mathrm{2D}$ & $M_r$, $C_\mathrm{2D}$& $M_i$, $C_\mathrm{2D}$ & $r_{50,\mathrm{2D}}$, $C_\mathrm{2D}$ \\
&  & 56.92($\pm$2.75) &56.25($\pm$2.37) & 56.20($\pm$2.56) & 55.91($\pm$2.58) & 55.75($\pm$2.37) \\ \cline{2-7}
&\multirow{2}*{3} &  $M_{\ast}$, $r_{90,\mathrm{2D}}$, $C_\mathrm{2D}$ & $M_{\ast}$, $r_{50,\mathrm{2D}}$, $C_\mathrm{2D}$& $M_i$, $r_{90,\mathrm{2D}}$, $C_\mathrm{2D}$ & $M_r$, $r_{90,\mathrm{2D}}$, $C_\mathrm{2D}$ & $M_i$, $C_\mathrm{2D}$, $r_{50,\mathrm{2D}}$\\
 & & 64.48($\pm$2.52) & 64.13($\pm$2.51) & 63.27($\pm$2.39) & 63.15($\pm$2.32) & 62.84($\pm$2.44)\\ \hline
\hline
\label{tbl:imp_combine_obs}
\end{tabular}
\end{center}
\end{table*}                  

The learning outcome and the feature importance rankings based on only observable features are practically more 
important, which we investigate in this subsection. In the right panel of Figure~\ref{fig:prediction}, the predicted 
values of $f_\mathrm{acc}$ are shown against their true values for the test sample when only observable galaxy features
calculated in projection are used for training and prediction. The $R^2$ score drops to 78.43\%, which is lower than 
the score (93.94\%) when all halo and galaxy features calculated in 3-dimensions are used (the left panel of Figure~\ref{fig:prediction}). In the bottom panel, the RMSE is on average 0.104, which is $\sim$40\% larger than the 
prediction by using all available halo and galaxy features calculated in 3-dimensions. There are also some biases 
from the diagonal line at small values of $f_\mathrm{acc}$.

Similar to Table~\ref{tbl:imprank}, the importance rankings for individual observable features are shown in 
Table~\ref{tbl:imprankobs}, for the full, high and low-mass samples. The scores and associated uncertainties are 
also presented in Figure~\ref{fig:impobs}.

All the features in Table~\ref{tbl:imprankobs} have smaller scores compared with corresponding features
in Table~\ref{tbl:imprank}, especially for $r_{90,\mathrm{2D}}$ and $r_{50,\mathrm{2D}}$. $r_{90,\mathrm{2D}}$
is no longer the most important individual feature for the full and low-mass samples. This is mainly because
$r_{90,\mathrm{2D}}$ is calculated based on mock galaxy images involving observational effects such as sky noise, 
PSF and dust attenuation. Projection plays a relatively minor effect. As we have explicitly checked, if $r_{90}$ 
is calculated by directly projecting the positions of star particles in the simulation, its importance can be as 
high as 80.74\% (full sample). The inclusion of sky noise has probably made part of the outer stellar halo drop 
below the noise level, and hence the determination of outer boundaries of galaxies becomes inaccurate
\citep[e.g.][]{2013ApJ...773...37H,2015MNRAS.454.4027D}, resulting in less correlation with the accreted stellar 
material. Nevertheless, $r_{90,\mathrm{2D}}$ is still the most important feature for the high-mass sample, 
while it remains in the top five most important features for the full sample and the low-mass sample. Similar to Table~\ref{tbl:imprank}, we find the ranking of $C_\mathrm{2D}$ is higher for the low-mass sample. 

Based on the same idea as Section~\ref{sec:impsim}, we also investigate the importance rankings of multiple observable feature combinations. The corresponding results are shown in Table~\ref{tbl:imp_combine_obs} for 
the full, high and low-mass samples. For the full sample, the most important two feature combination is
$M_{\ast}$ plus $r_{90,\mathrm{2D}}$. The other important two feature combinations are usually one stellar 
mass/luminosity or size related feature plus galaxy concentration, $C_\mathrm{2D}$, but the difference in the 
associated $R^2$ scores for the top five most important combinations is not significant compared with the errors. 
For the combination of three observable galaxy features, the scores are further increased. The top five most 
important combinations are usually one stellar mass/luminosity feature, one galaxy size feature plus $C_\mathrm{2D}$. 
We find $C_\mathrm{2D}$, a feature related to observed galaxy morphology in projection, appears very frequently, 
despite the fact that it is only the 5-th to 7-th most important individual feature in Table~\ref{tbl:imprankobs}.
This is because it carries the most amount of independent information from galaxy stellar mass and size. 

The most important two or three feature combinations for the high and low-mass samples are similar to those 
for the full sample. However, for the high-mass sample,the line-of-sight velocity dispersion, $\sigma_{z,\ast}$, 
sometimes appears to replace $M_\ast$, and the galaxy $g-r$ colour sometimes appears to replace $C_\mathrm{2D}$. 
Note $g-r$ is the least important feature in Table~\ref{tbl:imprankobs}. 
To summarize, a combination of up to three observable galaxy features of different types, galaxy stellar mass (or 
line-of-sight velocity dispersion), galaxy size plus a third morphology (or colour) related feature would lead to a 
score very close to the case when all observable galaxy features are used. In fact, using the combination of $M_\ast$,
$r_\mathrm{90,2D}$ and $C_\mathrm{2D}$ for the prediction, the RMSE is $\sim$0.119, which is similar to the RMSE 
of 0.104 for the right panel of Figure~\ref{fig:prediction}.

In the next subsection, we move on to directly compare the bias and scatter in the two cases, and compare 
with the prediction when only stellar mass is used as the input feature.


\subsection{Scatter and bias}
\label{sec:scatter}
\begin{figure*} 
\includegraphics[width=1.0\textwidth]{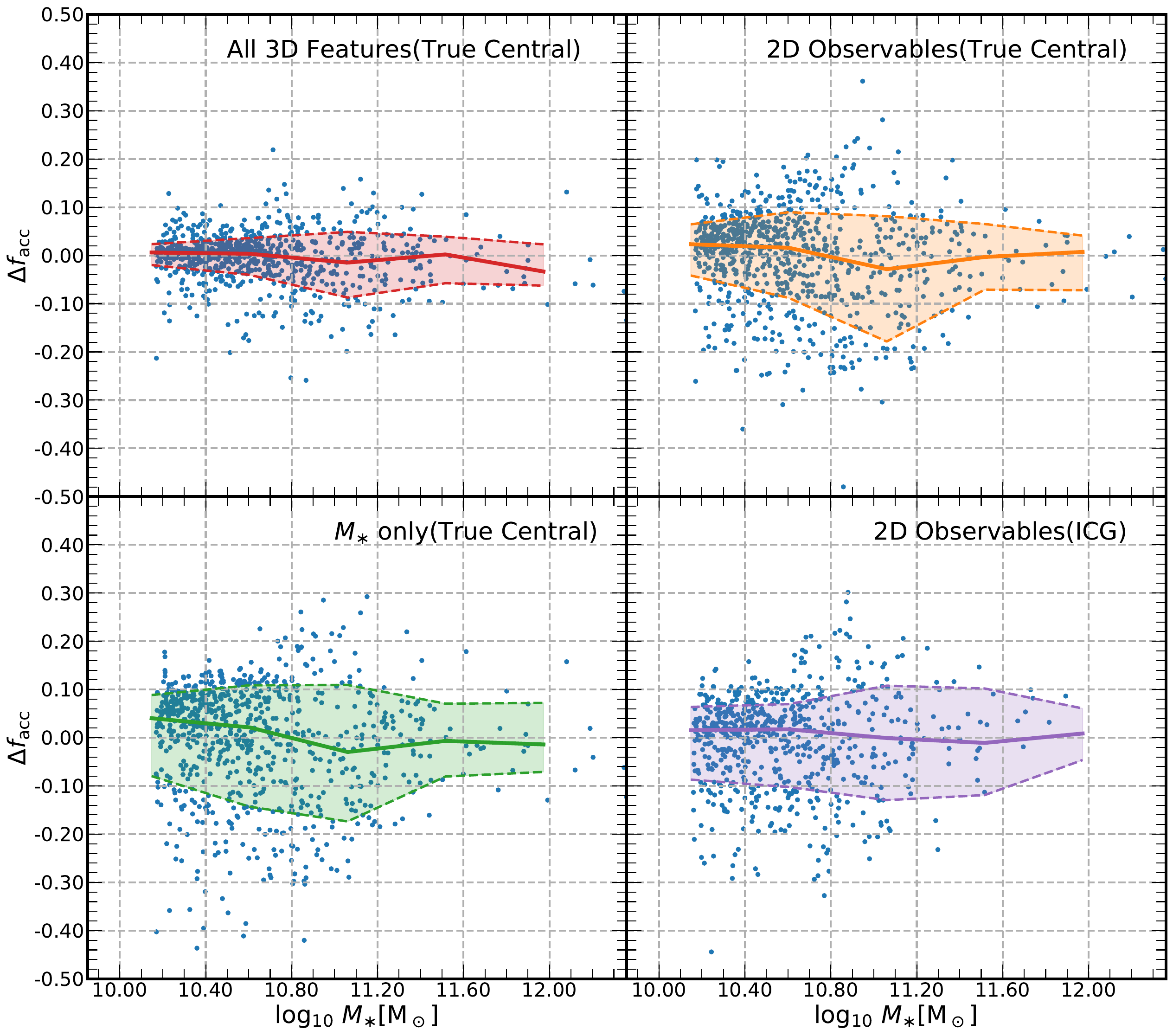}%
\caption{{\bf Upper Left:} The difference between predicted and true values of $f_{\mathrm{acc}}$ ($\Delta f_{\mathrm{acc}}$),
reported as a function of stellar mass. The prediction is based on the learning outcome when all available halo and galaxy
features calculated in 3-dimensions are used. Blue dots represent individual haloes/galaxies, while the solid and two dashed
lines show the median, 16-th and 84-th percentile boundaries, respectively. {\bf Upper Right:} The prediction is based on
observable galaxy features calculated in projection. {\bf Lower Left:} $f_{\mathrm{acc}}$ is predicted by only using the
stellar mass. {\bf Lower Right:} Similar to the top right panel, but is based on isolated central galaxies (ICG). Except for
the bottom right panel, the test sample used in all the other three panels are based on the same set of central galaxies from
TNG100-1. We take one test sample as representative. Other test samples after random divisions can show slightly different
results, but the general trends remain very similar. In addition, for all the four panels, the stellar mass in the $x$-axis is
the true stellar mass based on all bound star particles from the simulation, although different types of stellar mass are used
as input features for the RF training and prediction (see Section~\ref{sec:halogalprop} for details). This is for the sake of
fair comparisons.}
\label{fig:prediction2}
\end{figure*}
\begin{figure} 
\includegraphics[width=0.49\textwidth]{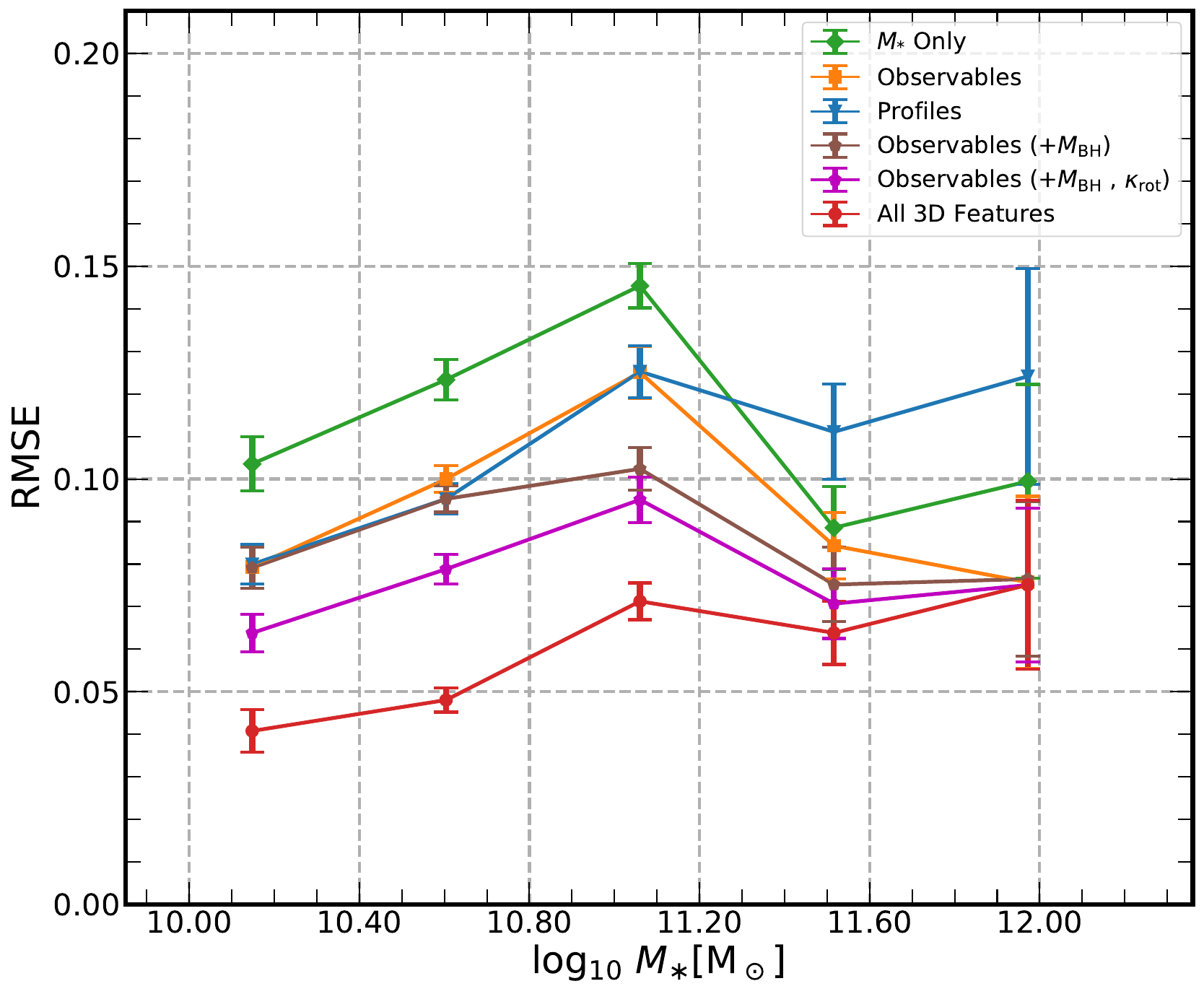}%
\caption{Red, orange and green curves show the root-mean-square errors versus stellar mass, which are based on 
the RF prediction trained using all available features calculated in 3-dimensions, using only observable features 
calculated in projection, and using the stellar mass only. These correspond to the upper left, upper 
right and lower left panels of Figure~\ref{fig:prediction2}, respectively. The blue curve is based on the RF 
prediction trained against surface brightness profiles in $gri$-bands, the projected stellar mass density profile, 
velocity and velocity dispersion profiles, density fluctuation at different radii and the axis ratio of the galaxy. 
The brown curve is for the prediction when the black hole mass is additionally included as a feature, and on the 
basis of the brown curve, the magenta curve further includes $\kappa_\mathrm{rot}$. Smaller values of RMSE mean 
better prediction. Errorbars are the 1-$\sigma$ scatters among the results of 100 randomly divided training and 
test samples. Small shifts along the $x$-axis have been manually added to all curves except the green one, in order 
to better display the errorbars. 
}
\label{fig:residual_analy}
\end{figure}

Figure~\ref{fig:prediction2} shows the residuals between the predicted and true values of $f_\mathrm{acc}$, 
i.e., $\Delta f_\mathrm{acc}$, which is reported as a function of stellar mass. Different panels refer to 
predictions based on different input features or different samples. 

The upper left panel shows the smallest amount of scatter, when all available halo and galaxy 
features calculated in 3-dimensions are included. Consistent with Figure~\ref{fig:prediction}, the scatter gets significantly 
larger in the upper right panel when only observable galaxy features are used. Despite the significant increase in the amount 
of scatter, the prediction remains almost unbiased, except for the low-mass end, where we see the data points tend to be skewed 
towards positive $\Delta f_{\rm acc}$\footnote{The amount of scatter and bias can be slightly decreased at the low-mass end 
if we use $\log f_\mathrm{acc}$ as the target instead of $f_\mathrm{acc}$. When using $\log f_\mathrm{acc}$ as the target 
variable, it is the bias and scatter in the ratio between predicted and true $f_\mathrm{acc}$ to be minimised, instead of the 
absolute difference.}.

Notably, Figure~\ref{fig:prediction2} is based on the learning outcome of the full sample. If we plot the results when the 
high and low-mass samples are trained separately, the scatter becomes slightly smaller at $\log_{10}M_\ast/\msun\sim11.2$, 
i.e., the mass threshold used for the division, but remains similar at the high and low-mass ends. Besides, we also note a 
mild peak in the amount of scatter at $\log_{10}M_\ast/\msun\sim11$, which is perhaps related to the transition in the relation 
between $f_\mathrm{acc}$ and stellar mass at $\log_{10}M_\ast/\msun\sim11$, as the readers can see from the bottom left panel 
of Figure~\ref{fig:correlation figure} that the scatter in $f_\mathrm{acc}$ seems to be slightly larger at 
$\log_{10}M_\ast/\msun\sim11$. Galaxies with $\log_{10}M_\ast/\msun\sim11$ represent a transition stage from blue star-forming 
disks to red passive ellipticals. Perhaps the modelling of these galaxies more strongly depends on a few other features,
which are so far not included in our analysis, such as the redshift when the star formation activity of the galaxy 
starts to quench. 

The difference between the two upper panels of Figure~\ref{fig:prediction2} reflects the fact that observable galaxy features 
are not enough to capture the full information of $f_\mathrm{acc}$. The prediction in the upper left panel is also based on a 
few unobservable features such as host halo mass, size, specific angular momentum and assembly histories. These have helped to 
reduce the scatter and bias significantly. In addition, other features used in the upper left panel, if observable, are in fact 
calculated in 3-dimensions without including observational noise and projection effects, which also improves the prediction.

Since the stellar mass is among one of the most important observable features, it is interesting to show the ability of predicting 
$f_\mathrm{acc}$ using stellar mass solely. This is shown in the lower left panel of Figure~\ref{fig:prediction2}. Compared with
the upper right panel, the amount of bias and scatter is slightly larger. At $\log_{10}M_\ast/\msun<11.1$, the distribution is 
a bit more asymmetric. Therefore the inclusion of a few other observable galaxy features indeed helps to bring some mild
improvements than simply using stellar mass. 

We also investigate whether the sample selection in real observation might affect the learning outcome. As mentioned in
Section~\ref{sec:icg}, we cannot directly identify true halo central galaxies in real observations, which are often selected 
through indirect or empirical approaches. Here we repeat our RF training against a sample of isolated central galaxies (see 
details in Section~\ref{sec:icg}), using only observable galaxy features calculated in projection. The scatter in predicted 
versus true values of $f_\mathrm{acc}$ based on the corresponding test sample is presented in the lower right panel of
Figure~\ref{fig:prediction2}. This is to be directly compared with the upper right panel. As have been mentioned, the selection 
of isolated central galaxies would introduce a small amount of contamination by satellite galaxies, but our results do not 
seem to be significantly affected by such satellite contamination, i.e., the scatter in the lower right panel seems to be even 
slightly smaller than that of the upper right panel. The slight decrease in the scatter of the lower right panel is perhaps 
due to the sample selection and statistical fluctuation.

Figure~\ref{fig:residual_analy} directly compares the three cases, by overplotting the RMSEs of the upper left, upper right 
and lower left panels of Figure~\ref{fig:prediction2} together. It is clearly shown that when only observable galaxy features 
are used, the RMSE of $f_\mathrm{acc}$ can be decreased by $\sim$0.03 (20\%) than purely using stellar mass for the prediction. 
When all available halo and galaxy features calculated in 3-dimensions are included, the scatter can be further reduced by 
$\sim$0.04. The amount of improvements is almost a constant at $\log_{10}M_\ast/\msun<11$, which is slightly smaller at
$\log_{10}M_\ast/\msun\geqslant 11$, perhaps indicating for high-mass galaxies, more information about $f_\mathrm{acc}$ 
is included in other features in addition to stellar mass and other unobservable features so far not included in our analysis. 
Consistent with Figure~\ref{fig:prediction2}, the RMSE itself also shows weak dependence on stellar mass, with a peak at
$\log_{10}M_\ast/\msun\sim11$, which slightly decreases towards both low and high-mass ends. Note when we only use the 
three most important features in combination to predict $f_\mathrm{acc}$ (all available features or observable features, see
Tables~\ref{tbl:imp_combine} and \ref{tbl:imp_combine_obs}) , the scatters are very similar to the red or orange curves in
Figure~\ref{fig:residual_analy}.

So far, after including all observable galaxy features in the RF training for prediction, the RMSE is about 0.1 in 
$f_\mathrm{acc}$, which gently decreases to $\sim0.08$ at the low and high-mass ends. One interesting question to ask is 
what is the best we can achieve, if we input entire galaxy images or velocity maps for training and prediction? We move 
on to investigate this in the next section. 

\section{Discussion}
\label{sec:disc}
\subsection{Using density and velocity profiles for prediction}
\begin{table*}
\caption{
Feature importances ($R^2$ scores) for the projected stellar mass densities, line-of-sight 
velocity and velocity dispersions measured at different radii. We do not provide results 
for surface brightness profiles in $gri$-bands, which actually show similar trends as the 
projected stellar mass density profile. The profiles are calculated for 10 radial bins between 
0.31 and 3.16 times the projected radius containing half of the stellar mass ($r_{50}$).}
\begin{center}
\begin{tabular}{lrrrrrrrrrrrrrrrrrrrr}\hline
\hline
&&\multicolumn{7}{c}{Importance (\%)}\\
$r/r_{50}$ & 0.36 & 0.45 & 0.57 & 0.71 & 0.90 & 1.13 & 1.42 & 1.79 & 2.25 & 2.84 \\ \hline
velocity dispersion profile & 44.57 & 41.81 & 38.95 & 35.67 & 31.44 & 27.24 & 22.24 & 19.30 & 17.77 & 16.84\\
density profile & 16.88 & 19.94 & 21.66 & 21.91 & 21.50 & 20.89 & 19.51 & 16.71 & 12.52 & 8.07\\
velocity profile & 5.82 & 5.33 & 5.12 & 4.77 & 4.22 & 3.36 & 2.84 & 2.28 & 2.13 & 2.07 \\
\hline
\label{tbl:imprankprofile}
\end{tabular}
\end{center}
\end{table*}  
\begin{figure} 
\includegraphics[width=0.49\textwidth]{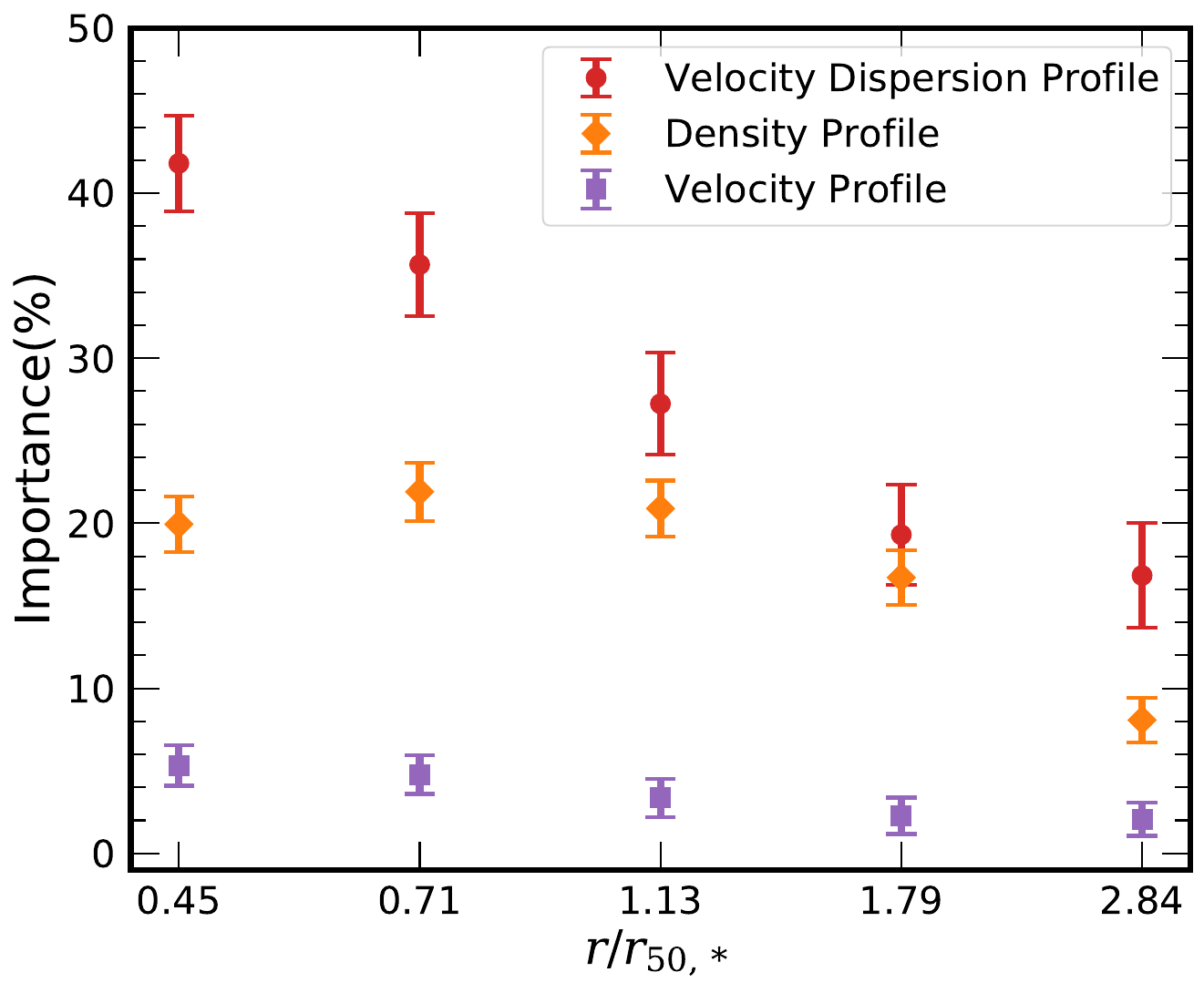}%
\caption{The feature importances for the projected stellar mass densities, line-of-sight velocities and velocity 
dispersions measured at different radii (red dots, orange diamonds and purple squares, see the legend). Errorbars 
are the 1-$\sigma$ scatters among the results of 100 randomly divided training and test samples.}
\label{fig:impprofile}
\end{figure}

In this subsection, we try to test what is the best one can achieve to predict $f_\mathrm{acc}$ with current and future 
observations. In principle, we can input the entire galaxy image in all available filters for training. We can also include 
the velocity map based on current or future integral field unit (IFU) observations. The images and velocity maps themselves 
should contain the maximum amount of information so far we can observe for a given galaxy. Since the training based on entire 
galaxy images is not efficient with the RF method, we choose to represent the image using the surface brightness profiles 
in $gri$-bands and the projected stellar mass density profile. The velocity map is represented by the mean line-of-sight 
velocity and velocity dispersion profiles, and we choose the $z$-axis of the simulation to represent the line-of-sight 
direction. We use these profiles as input features. We also postpone analysis based on galaxy images and velocity maps
to future studies with alternative machine learning algorithms such as the neural network.

To calculate the various kinds of profiles, we adopt 10 radial annuli between 0.31 and 3.16 times the projected 
radius containing half of the stellar mass. We have explicitly checked that our RF learning outcome and feature 
importance ranking are not sensitive to how the binning is chosen, as long as the number of bins is in a reasonable
range. Moreover, TNG galaxies are not strictly spherical, and for a given radial annulus, the actual density and 
surface brightness distribution can vary within the annulus. Thus we also include the axis ratio of the entire galaxy 
and the density fluctuations at different radii as input features. However, the scores for these features are low, which 
help little to improve the prediction, and thus we do not directly show their importances. We have also tried to include 
the covariance matrixes for all these profiles as input features, and we end up with almost no improvements.

Blue lines and symbols in Figure~\ref{fig:residual_analy} show the RMSEs of the learning outcome based on all the 
different kinds of profiles mentioned above, as a function of stellar mass. Compared with the orange lines and symbols, 
which have been discussed in Section~\ref{sec:scatter}, the blue lines and symbols have slightly smaller values at 
$\log_{10}M_\ast/\msun\sim10.6$, whereas the RMSEs in the two most massive bins even become larger though the errors 
are also quite large. Therefore, it seems even after we use up all the information contained in the projected density, 
surface brightness, line-of-sight velocity and velocity dispersion profiles, very limited improvements can be achieved. 
This is probably because for mergers/accretions happened early on, stars have enough time to be phase mixed, leaving 
less amount of information about the merger histories in the density and velocity maps. Maybe further including tangential
velocities and metallicity gradients can help to achieve more improvements, though tangential velocities can be extremely
difficult or impossible to measure for distant extra-galactic systems. In addition, satellite galaxies can continue forming
stars after falling into the current host halo. The amount of stellar mass formed after infall also contributes to the 
ex-situ stellar mass. Further including features related to the star formation activities and orbits of satellites after 
infall might help to improve. 

The feature importances for the projected stellar mass density profile, the line-of-sight velocity and velocity dispersion profiles 
measured at different projected radii are presented in Table~\ref{tbl:imprankprofile} and Figure~\ref{fig:impprofile}. We only plot 
5 out of the 10 radial bins in Figure~\ref{fig:impprofile} to make the figure easy to read. In Figure~\ref{fig:impprofile}, the red 
dots are all above the purple squares, especially for those points in more central regions, indicating that the second moments 
in velocity carry significantly more information than the first moments. Besides, it seems the projected stellar mass 
density profiles are significantly less important than the second moments in velocity as well, but are more important than the 
first moments in velocity, with the orange diamonds in the middle of red dots and purple squares. 

Indeed, substructures formed by mergers are expected to preserve better their clustering in the full 6-dimensional phase space. 
As a result, many previous studies look for tidal debris and substructures around our Milky Way Galaxy and also extra-galactic 
galaxies with IFU observations in velocity, energy and action spaces \citep[e.g.][]{2004ApJ...610L..97H,2018ApJ...856L..26M,2018MNRAS.478.5449M,
2018MNRAS.478..611B,2020ApJ...891...39Y,2021ApJ...919..135D,2021arXiv211013172Z}. The fact that we found the second moments in 
the velocity are more important than the density distribution is consistent with the expectation, proving that numerical simulations
performed under the standard hierarchical structure formation theory of our Universe is capable of producing such a trend. Moreover,
Figure~\ref{fig:impprofile} also shows that the importance of the velocity dispersion profiles shows a dependence on the radius, 
i.e., the velocity dispersion in the central region is more important. This is perhaps because tidal debris is more phase mixed in
central regions of their hosts \citep[e.g.][]{2015MNRAS.453..377W}, and thus velocity dispersion information becomes more important 
to disentangle the in-situ and ex-situ components.  

The fact that even after including the entire projected density, surface brightness, line-of-sight velocity and velocity dispersion 
profiles, the scatter in the learning outcome is not significantly improved, indicates that multi-component decomposition of the 
surface brightness profiles would suffer from similar or even worse amount of uncertainties, compared with the orange or blue solid 
lines in Figure~\ref{fig:residual_analy}. 

\subsection{Improvements with future observations}

So far we did not include the mass of central black holes, $M_\mathrm{BH}$, in our list of features. Observations have revealed 
tight correlations between the mass of central supermassive black holes and properties of host galaxies, such as the velocity 
dispersion, mass and luminosity of the bulge component, total stellar mass and velocity dispersion of the host \citep[see e.g.][
for a review]{2013ARA&A..51..511K}. These relations indicate strong co-evolution of central black holes and their host galaxies. 
In many previous studies, this is often interpreted as the outcome of galaxy mergers, which contribute to the growth in mass of
both central black holes and host galaxies 
\citep[e.g.][]{2006MNRAS.365...11C,2006ApJ...652..864H,2007ApJ...671.1098P,2011Natur.469..377K,2021arXiv211203576M}, 
though many other studies also proposed merger-free scenarios to interpret the co-evolution
\citep[e.g.][]{2004ARA&A..42..603K,2010ApJ...721...26G,2012ApJS..198....4O,2013MNRAS.429.2199S}.

According to the above background scenario, $f_\mathrm{acc}$ might be correlated with $M_\mathrm{BH}$. 
Observationally, $M_\mathrm{BH}$ is 
difficult but still possible to measure, through, for example, reverberation mapping of AGN broad line regions
\citep[e.g.][]{1982ApJ...255..419B}. We thus examine the learning outcome when $M_\mathrm{BH}$ is used as an observable. 
The brown lines and symbols in Figure~\ref{fig:residual_analy} show the RMSE after further including $M_\mathrm{BH}$ as an 
observable feature. Compared with the blue lines and symbols, the RMSE is further reduced by at least $\sim$20\% at $\log_{10}M_\ast/\msun>11$, but the error of the most massive data point is very large. There are very weak improvements 
at the low-mass end. If $M_\mathrm{BH}$ can be accurately measured in future observations, the prediction of $f_\mathrm{acc}$ 
by additionally including $M_\mathrm{BH}$ is promising to be further improved by at least $\sim$20\% for massive galaxies, 
though this conclusion might depend on the subgrid physics, as TNG is not capable of resolving black hole mergers. 

In fact, as we have checked, in TNG100, $M_\mathrm{BH}$ is very strongly correlated with the host halo mass defined 
through all bound particles, $M_\mathrm{halo}$,  with a considerably smaller amount of scatter than that between $M_\ast$ 
and $M_\mathrm{halo}$\footnote{In TNG100, the amount of scatter between $M_\mathrm{BH}$ and $M_\mathrm{halo}$ is also 
considerably smaller than that between the total stellar mass in satellites with $\log_{10}M_\ast/\msun>8$ and $M_\mathrm{halo}$.}. 
Thus the inclusion of $M_\mathrm{BH}$ is a very good reflection of the host halo mass for central galaxies in TNG, 
resulting in such a significant improvement in the prediction of $f_\mathrm{acc}$ for massive galaxies. However, also 
due to this strong correlation, if further including $M_\mathrm{BH}$ in the list of all halo and galaxy features calculated 
in 3-dimensions, the improvement is weak, i.e., the RMSE is similar to the red lines and symbols in 
Figure~\ref{fig:residual_analy}.

Many existing studies have used the abundance, total luminosity or total stellar mass of satellite galaxies as a proxy 
to the host halo mass \citep[e.g.][]{2012MNRAS.424.2574W,2013MNRAS.428..573S,2019arXiv191005139G,2021ApJ...919...25W}. We 
did not include these satellite related features in our analysis due to the concern of the TNG100 resolution limit. 
Besides, in current spectroscopic observations, it is difficult to measure the satellite abundance down to faint magnitudes. 
Many small satellites do not have spectroscopic observations. The method of counting and averaging photometric satellites 
around spectroscopically identified central galaxies, with statistical background subtraction or modelling of photometric 
redshift errors \citep[e.g.][]{2011ApJ...734...88W,2012MNRAS.427..428G,2016MNRAS.459.3998L,2019ApJ...879...71W,
2021SCPMA..6489811W,2021arXiv211005760X}, cannot achieve satellite counts for individual systems. However, on-going and 
future deep spectroscopic surveys such as the Dark Energy Spectroscopic Instrument \citep[DESI;][]{2016arXiv161100036D}, 
will provide spectroscopic measurements for much fainter satellites. We thus further check the amount of improvements 
after including the total stellar mass in satellites more massive than $10^8\msun$ as an observable feature\footnote{Here 
for central galaxies without any satellite above this threshold, we simply exclude them.}. Unfortunately, the improvement 
is very limited, which is at most $\sim$1\% at $\log_{10}M_\ast/\msun\sim11.1$. In particular, the corresponding score is 
52.19\% for the full sample, and the small improvement is mainly due to its correlation with $M_\ast$ in TNG. 

In addition to $M_\mathrm{BH}$ and satellites, another feature, $\kappa_\mathrm{rot}$, which is not considered as an 
observable in our analysis above, is in fact closely related to the so-called spin parameter of galaxies. The spin 
parameter quantifies the specific angular momentum, which can be estimated for galaxies with good IFU data
\citep[e.g.][]{2016ARA&A..54..597C,2019arXiv191005139G,2020MNRAS.495.1958W,2021arXiv211013172Z}. The magenta lines 
and symbols in Figure~\ref{fig:residual_analy} show the RMSE after further including $\kappa_\mathrm{rot}$ as an 
observable, together with $M_\mathrm{BH}$. The inclusion of $\kappa_\mathrm{rot}$ further decreases the RMSE for 
smaller galaxies. Since $M_\mathrm{BH}$ is more important for massive galaxies, the joint inclusion of $M_\mathrm{BH}$ 
and $\kappa_\mathrm{rot}$ help to decrease the RMSE by $\sim$20\% over the entire mass range. Note our analysis is 
based on the true values of $M_\mathrm{BH}$ and $\kappa_\mathrm{rot}$ from the simulation, without any observation 
and projection effects. So the 20\% of improvement should be understood as an estimate of the upper limit for what 
we can achieve, if $M_\mathrm{BH}$ and $\kappa_\mathrm{rot}$ can be accurately determined with future observations. 

The fact that $r_{90,\mathrm{3D}}$ is the most important feature is very promising. Our results show that 
after including the sky noise and PSF of SDSS, the importance of $r_{90,\mathrm{2D}}$ drops. However, on-going 
and future deep imaging surveys and instruments, such as the Hyper Suprime-Cam (HSC) Imaging Survey 
\citep{2012SPIE.8446E..0ZM,2018PASJ...70S...1M,2018PASJ...70S...2K,2018PASJ...70S...3F},
the Large Synoptic Survey Telescope \citep[LSST;][]{2008arXiv0805.2366I} and the China Space Station Telescope 
\citep[CSST;][]{zhan2011,Cao2018,Gong2019} are promising to achieve significantly better resolution and deeper 
imaging of the outer stellar halo, hence improving the measurement of $r_{90,\mathrm{2D}}$ and the prediction of 
$f_\mathrm{acc}$.

\subsection{Can the learning outcome be directly applied to real galaxies?}
\label{sec:tension}

In this subsection, we further discuss the feasibility and uncertainty of applying our RF learning outcome based 
on TNG100-1 galaxies to real galaxies. We have shown that the inclusion of other observable galaxy features in 
addition to stellar mass can lead to $\sim$0.03 (20\%) of decrease in the RMSE, so the RF prediction based on all 
these features jointly is indeed more precise than simply predicting $f_\mathrm{acc}$ based on only stellar mass.
Moreover, even after including almost all available information from galaxy images and velocity maps, the uncertainty 
in the prediction of $f_\mathrm{acc}$ is about 0.1, which gently decreases to $\sim0.08$ at the low and high-mass 
ends. This is perhaps the limit one can achieve with current observations. 

However, to apply the RF learning outcome based on TNG galaxies to real observed galaxies, we care about how TNG 
agrees with the real data. Disagreement between TNG predictions and the real Universe would introduce systematic 
uncertainties. It has been shown in previous studies that TNG predictions still deviate from real data, in terms 
of a few detailed and stringent comparisons, such as the mass-size relation, morphological and colour features.

For example, by comparing the mass-size relation of SDSS and GAMA galaxies with TNG predictions, \cite{Pillepich2018}
showed that TNG galaxies tend to have larger sizes than those of real galaxies at fixed stellar mass. Moreover,
\cite{Genel2018} quantitatively estimated that the discrepancies between the sizes of TNG100 and real galaxies are 
in the range of $\sim$0 to 0.2dex, which is sensitive to sample selections and size definitions in both observations 
and simulations.

In addition to the mass-size relation, tensions also exist when comparing the morphology of TNG and observed 
galaxies. For example, \cite{Rodriguez-Gomez2019} compared synthetic galaxy images from TNG and Pan-STARRS observations.
It was reported that TNG galaxies show good overall agreement with real data. The median trends of morphological, 
size and shape features with respect to stellar mass are consistent with the observational trends within 1-$\sigma$. 
However, TNG has difficulties in reproducing a strong morphology-colour relation, which is probably due to inefficient 
feedback near the galaxy centre, and thus TNG galaxies are too concentrated. With deep learning classifications, the 
mass-size relations of TNG galaxies divided by morphological types agree well with the relations of real galaxies, 
but the correlation between optical morphology and S\'{e}rsic index is weaker than SDSS \citep{2019MNRAS.489.1859H}. 
Moreover, \cite{Zanisi2021} announced that TNG simulations still cannot reproduce the details of galaxy morphology 
structures on small scales very well, especially for those quenched spheroidal galaxies, which could be related to the 
coarse numerical resolution.

Delicated disagreement also exists in the outer stellar haloes and host dark matter haloes between TNG and real 
observed galaxies. \cite{2020MNRAS.495.4570M} compared a few galaxies from the Dragonfly Nearby Galaxies Survey with 
TNG mass-matched counterparts, and found that real galaxies have less mass or light at large radii, which they denote 
as a so-called ``missing outskirts problem". The inconsistency might be related to their small sample of nearby galaxies, 
which is not large enough to achieve good statistical significance, but it might also indicate the necessity of a finer 
tuning in satellite disruption in TNG. More recently and combining weak lensing measurements, \cite{2021MNRAS.500..432A}
reported that at the virial mass of $\sim 10^{13}\msun$, the outer stellar mass profile of massive galaxies at $z\sim0.4$ 
is in excellent agreement with TNG predictions, whereas at $\sim 10^{14}\msun$, TNG shows excess in the outer stellar mass,
though the effect of extended PSF wings \citep[e.g.][]{2019MNRAS.487.1580W} is not carefully corrected. Moreover,
\cite{2020MNRAS.498.5804R} claimed that TNG300 tends to predict $\sim$50\% more lensing signals for red galaxies with
$10.2<\log_{10}M_\ast/\msun<11.1$ at a projected distance of 0.6Mpc/h to the galaxy centre. 

At the massive end, the stellar mass and luminosity functions of simulated and observed galaxies often 
show discrepancies, which might be due to the missing outskirts of observed galaxies below the sky noise 
level of shallow surveys \citep[e.g.][]{2013ApJ...773...37H,2015MNRAS.454.4027D,2018MNRAS.475.3348H}. Such 
a discrepancy can be significantly reduced after adopting some aperture cuts to simulated galaxies 
\citep{Pillepich2018}, but does not totally disappear.

According to the studies mentioned above, further improvements are still required to bring better agreement 
between TNG predictions and real observations, especially in terms of galaxy size and morphology. In our analysis
of this paper, we found up to three features related to stellar mass, galaxy size and morphology capture the most 
amount of information upon determining $f_\mathrm{acc}$. Since TNG galaxy size and morphology still show mismatches 
with observations, applying our current RF model directly to real data can be problematic. Note, however, the 
tension with real observation is not a unique problem to only TNG, but stays as a challenge to many other hydrodynamical 
simulations as well. For instance, the galaxy sizes predicted by the EAGLE \citep[Evolution and Assembly of GaLaxies 
and their Environments simulations;][]{Schaye2015,Crain2015} suite of simulations were also reported to be larger 
than real observed galaxies \citep[e.g.][]{vandeSande2019,2021arXiv211004434Y,2022MNRAS.511.2544D}.
Moreover, \cite{Bignone2020} reported that the distribution of non-parametric morphological statistics based on mock 
and real observed galaxy images still present non-negligible differences, which is true for both TNG and EAGLE. 

As a result, efforts are still necessary to bring in better agreements between predictions by 
modern hydrodynamical simulatsions and real data, before we can finally apply our RF model to real observations. 
However, our results do suggest that a joint modelling of all or up to three different observable features based 
on the RF machine learning approach can lead to a precision of about 0.1 in the prediction of $f_\mathrm{acc}$, 
and if the central black hole mass and spin parameter of galaxies can be accurately measured in future observations, 
the uncertainty can be further decreased by $\sim$20\%. Hence the RF learning outcome is promising to be applied 
to future simulations and observations. Notably, we have repeated our analysis with the higher resolution TNG50
simulations. Though the numbers of galaxies in the training and test samples significantly decrease due to 
the smaller box size, our conclusions about the feature importance rankings remain very similar.

\section{Conclusions}
\label{sec:concl}

Using the TNG100-1 simulation and the random forest (RF) machine learning approach, we have performed 
a comprehensive study on the importance ranking of various halo/galaxy features and the prediction of 
the ex-situ stellar mass fractions ($f_\mathrm{acc}$) for galaxies with $\log_{10}M_\ast/\msun>10.16$.
The amount of bias and scatter in the learning outcome is carefully investigated.

The default feature importance ranking returned by the RF method suffers from the so-called masking effect 
when there are strong correlations among input features. We thus use the $R^2$ score based on each individual 
feature and different feature combinations to quantify the importance ranking.

We find for high-mass galaxies within $\log_{10}M_\ast/\msun>11.20$, global halo and galaxy features, 
including the virial mass of the host halo, the total mass of all bound particles, the virial radius of the host 
halo and the stellar mass of the galaxy, are among the few most important features. For low-mass galaxies with
$10.16<\log_{10}M_\ast/\msun \leqslant 11.20$, the importances for features reflecting halo assembly histories
and galaxy morphologies are increased. This is very probably because the star formation of low-mass galaxies is 
dominated less by the shallower potential of the host halo, and could be more stochastic and sensitive to merger 
histories. 

For the full, low and high-mass samples of galaxies used in our analysis, we found the radius containing 90\% of 
the total stellar mass, calculated in 3-dimensional spherical shells, $r_{90,\mathrm{3D}}$, always has the highest 
importance. This is probably because $r_{90,\mathrm{3D}}$ defined in this way is close to the boundaries of galaxies 
and probes well the mass accreted and deposited in the outskirts of galaxies. However, if it is calculated from 
mock galaxy images in projection, after involving dust attenuation, the PSF and sky noise from SDSS, the importance 
of $r_{90,\mathrm{2D}}$ decreases, though it is still among the few most important observable features. This is mainly 
caused by the inclusion of sky noise and other observational effects, while projection plays a minor role. Compared 
with $r_{90,\mathrm{3D}}$, the importance of $r_{50,\mathrm{3D}}$ as an individual feature is significantly lower, 
indicating it is the outer stellar halo more strongly correlated with mergers. For the full, high and low-mass samples, 
the $g-r$ colour, stellar age and star formation rate are all not important individually. 

We find a combination of up to three features with different types can already saturate the predictive power. 
Theoretically, the combination of one global feature related to halo/galaxy mass or size, another feature reflecting 
the halo assembly history and a third morphological feature such as galaxy concentration, can lead to scores very close 
to the case when all available halo and galaxy features calculated in 3-dimensions are used. Observationally, the 
combination of galaxy stellar mass, galaxy size and a third feature reflecting galaxy morphology can lead to scores 
very similar to the case when all observable features are used. For more massive galaxies, stellar mass can be replaced 
by the line-of-sight velocity dispersion, and the morphological feature can be replaced by galaxy $g-r$ colour.

When all available halo and galaxy features calculated in 3-dimensions are used, the RF learning outcome gives a small 
amount of scatter, with an average RMSE of 0.068 in $f_\mathrm{acc}$. If using the combination of three features
quantifying galaxy size, assembly history and morphology, the RMSE is already as small as 0.072. On the other hand, when 
only observable galaxy features calculated in projection are used, the bias is still negligible, except for the low-mass 
end. However, the scatter is significantly increased, with the average RMSE of $f_\mathrm{acc}$ increasing to 0.104, which 
shows only weak dependence on stellar mass. Nevertheless, if compared with the case when only stellar mass is used for 
training and prediction, the inclusion of other observable galaxy features can indeed help to decrease the amount of 
scatter by $\sim$20\%. If using the combination of three observable features quantifying galaxy stellar mass, size 
and morphology, the RMSE is already as small as 0.119.

To understand what is the best we can achieve with current observations, we tried to use the entire projected stellar mass 
density profile, surface brightness profiles in $gri$-bands, the velocity and velocity dispersion profiles, the density 
fluctuation at different radii and the axis ratio of the galaxy as input features. We found velocity dispersions are more 
important than the surface density profiles, while the mean velocities are the least important. It turns out that the scatter 
in the learning outcome is only slightly improved by $\sim$5\% at $\log_{10}M_\ast/\msun<11.1$ after including all these
profiles. The profiles included in this way are expected to represent the entire galaxy image and velocity map, and thus 
carry approximately the maximum amount of information that we can observe. The limited amount of improvements perhaps 
implies that the merger histories cannot be fully captured by the density and velocity maps, especially for stars merged 
a long time ago and are more completely phase mixed with the in-situ component. 

With current observations, it is thus difficult to constrain $f_\mathrm{acc}$ better than a precision of $\sim$0.1. Our
findings also indicate that multi-component decomposition based on the surface brightness profiles galaxies should suffer 
from at least a similar or even worse amount of uncertainties than our RF approach. 

We find the inclusion of the central black hole mass as an observable can further decrease the scatter for massive galaxies.
If additionally including the spin parameter of the host galaxies as an observable, the scatter can be improved for smaller
galaxies. The central black hole mass is very strongly correlated with the host halo mass in TNG100. In the future, if the 
central black hole mass and the spin parameter of galaxies can be more accurately determined in observation, the scatter is
promising to be further reduced by $\sim$20\%.

\section*{Acknowledgements}

This work is supported by NSFC (12022307, 11973032, 11833005, 11890691, 11890692, 11621303, 12125301, 11773001, 12192222), National Key Basic Research and Development Program of China (No.2018YFA0404504), 111 project (No.
B20019), Shanghai Natural Science Foundation (No. 19ZR1466800) and the science research grants from the China 
Manned Space Project (No. CMS-CSST-2021-A02, CMS-CSST-2021-B03, CMS-CSST-2021-A03, CMS-CSST-2021-A07). We 
gratefully acknowledge the support of the Key Laboratory for Particle Physics, Astrophysics and Cosmology, 
Ministry of Education. We thank the sponsorship from Yangyang Development Fund.

We thank the anonymous referee for his/her useful comments and careful reading of this paper.
We are grateful for useful discussions on galaxy formation physics with Caina Hao, Lizhi Xie and 
Xiaohu Yang. We thank helpful discussions on details about the Eagle simulation with Shi Shao.
RS is grateful for discussions on the random forest machine learning method and algorithm with Yanrui 
Zhou and Wei Zhang, on halo angular momentum with Yifeng Zhou and Fuyu Dong and on galaxy images with 
Xiaokai Chen and Chengze Liu. RS is also grateful for the help by Ziyang Chen to download the TNG 
dataset. 

This work has made extensive use of the \textsc{python} packages -- \textsc{ipython} with its \textsc{jupyter} 
notebook \citep{ipython}, \textsc{numpy} \citep{NumPy} and \textsc{scipy} \citep{Scipya,Scipyb}. All the figures 
in this paper are plotted using the python matplotlib package \citep{Matplotlib}. This research has made use of 
NASA's Astrophysics Data System and the arXiv preprint server. The computation of this work is carried out on 
the \textsc{Gravity} supercomputer at the Department of Astronomy, Shanghai Jiao Tong University.

\section*{Data availability}
Data available to share upon request.
\bibliography{master}

\end{document}